# LLM4SCREENLIT: Recommendations on assessing the performance of large language models for screening literature in systematic reviews

Lech Madeyski [a],*, Barbara Kitchenham [b], Martin Shepperd [c]

[a] Wroclaw University of Science and Technology, Wyb. Wyspianskiego 27, Wroclaw, 50-370, Poland
[b] Keele University, UK
[c] Brunel University of London, UK

ABSTRACT

**Context:** Large language models (LLMs) are increasingly used to screen literature for systematic reviews (SRs), but the standard confusion-matrix metrics used to evaluate them can mislead under the imbalanced, cost-asymmetric conditions of screening.
**Objective:** We develop and justify LLM4SCREENLIT — practical recommendations for researchers conducting LLM-screening evaluations and for editors and reviewers assessing such studies — differentiated by study type (retrospective benchmarking vs. deployment for a specific SR).
**Method:** Using Delgado-Chaves et al. (2025), an 18-LLM benchmark across three biomedical SRs, as a motivating example, we reviewed 28 additional papers and extracted their reported metrics. We propose a Weighted Matthews Correlation Coefficient (WMCC) that integrates MCC's chance-correction with asymmetric misclassification costs, and validated it on three software-engineering (SE) reanalyses (Felizardo et al. 2024; Syriani et al. 2024; Huotala et al. 2025), the largest covering 9 LLMs × 24 SE secondary studies (34,528 articles).
**Results:** Across the 29 papers, only 10% reported MCC, only 24% reported full confusion matrices, and none of the five papers claiming workload savings priced false-negative cost. In the largest SE reanalysis, MCC and WMCC disagree on the best LLM in 55% of evaluable studies; in the most striking 9695-article SE study, the Accuracy-best LLM loses 63.3% of relevant evidence (Lost Evidence), the MCC-best 43.9%, but the WMCC-best only 5.8%. Sensitivity analysis (median crossover at $w \approx 2.7$, all $< 7$) supports $w = 10$ as a conservative default.
**Conclusions:** SR-screening evaluations should prioritise Lost Evidence and use cost-sensitive WMCC alongside MCC for ranking. Reporting must include the full confusion matrix and treat unclassifiable outputs as positives requiring human review. Designs should be leakage-aware, with non-LLM baselines when the study aims to inform SR practice and labels are available. Editors and reviewers should require these elements as routine. Extension to full-text screening and data extraction is principled but pending empirical validation.

## 1. Introduction

Large language models (LLMs) are increasingly employed to automate the challenging task of paper screening in systematic reviews (SRs),[1] promising substantial reductions in human workload and faster evidence synthesis in software engineering [1–6] and beyond [7–14]. To quantify their effectiveness, standard confusion-matrix metrics, e.g., accuracy, precision, recall, specificity, and F1-score, are adopted by comparing model decisions (include/exclude) against human reference labels. Although these metrics offer a familiar evaluation framework, their uncritical application can yield misleading conclusions. In particular, we argue it is essential to consider the features of the problem domain and which metrics best address them. This contrasts with an evaluation using all the 'usual' metrics in the hope that somehow useful insights might emerge.

We argue that there are four key features relating to screening papers for SRs. First, the data will tend to be extremely imbalanced so that the negative class (i.e, papers that are irrelevant to the SR) will considerably outnumber the positive class (i.e., relevant papers). Second, the costs of misclassifications are unequal. A relevant paper wrongly excluded, i.e., a false negative (FN), will likely have a far greater impact upon the quality of the SR than an irrelevant paper that passes the screening, i.e., a false positive (FP), and then wastes human effort subsequently rectifying the situation. Third, resources are limited, so we are concerned about the overall costs and benefits of deploying different screening tools. Fourth, we would like to be reassured that sophisticated, yet essentially black-box methods such as LLM screening tools are actually doing better than guessing.

* Corresponding author.
*E-mail address:* lech.madeyski@pwr.edu.pl (L. Madeyski).

[1] Systematic reviews are also referred to as systematic literature reviews or SLRs in the software engineering literature. We use SR throughout this paper but retain SLR where cited studies use it as part of their own terminology (e.g., Felizardo et al.'s dataset names SLR1 and SLR2).

While the studies we review span multiple domains, including biomedicine, the methodological challenges we address — class imbalance, asymmetric misclassification costs, and the need for chance-anchored metrics — are fundamentally mathematical rather than domain-specific. These recommendations are directly relevant to the growing adoption of LLMs for SR screening in SE [1,3–6].

Delgado-Chaves et al. [7] (hereafter referred to as DC+) recently evaluated 18 LLMs for screening studies in three SRs covering physiotherapy, neurology, and digital health topics (hereafter referred to as SR-I, SR-II, and SR-III), comparing selections with human reviewers and using confusion matrix metrics. They also emphasise comparing LLMs with more traditional machine learning methods, specifically the random forest method. While DC+ represents a valuable and ambitious contribution to assessing LLMs for SR screening, we use it as an illustrative example to highlight methodological issues that are common across the fields — such as reporting Accuracy as a primary metric under class imbalance — because DC+ provides complete confusion matrices that enable detailed reanalysis. Note that despite being drawn from the biomedical domain, we have used DC+ as our motivating example because it, in some regards, represents good practice, is published in a prestigious and influential venue, and because many pioneering and important methodological advances have come originally from this domain and then have subsequently been adopted by fields such as software engineering.

We also report the results of reviewing 28 other papers that studied the use of LLMs to screen literature for SRs. The papers were obtained from the primary studies included in two systematic reviews [15,16], together with papers we found from informal searches at the beginning of our investigation. We investigated these papers for three purposes: (i) to confirm that the problems we observed in the DC+ paper are not unique to that paper, (ii) to assess whether any other problems exist, and (iii) to investigate whether there are additional good practices to recommend.

The remainder of the paper is organised as follows. First, in Section 2 we describe our motivating study DC+ and frame it in the context of confusion matrices and how these provide a useful abstraction. From here we proceed (in Section 3) to the current, wider literature on LLM-based screening of studies for SRs, including SE-specific worked examples that illustrate the impact of metric choice on SE evaluations (Section 3.4). From these two sources, we then (Section 4) derive the LLM4SCREENLIT recommendations and implications, structured by study type (benchmarking vs. deployment), and conclude with some limitations of the study and suggestions for further work.

## 2. Issue with confusion matrix metrics

This section explains why correctness inadequately measures LLM performance in SRs, highlighting Recall and Lost Evidence as critical metrics for literature screening. Throughout this paper, we refer to the counts from confusion matrices as True Positives (TPs), True Negatives (TNs), False Positives (FPs), and False Negatives (FNs). The formulas used to construct the confusion matrix metrics discussed in this section can be found in Appendix A.

### 2.1. The fallacy of correctness

The DC+ study abstract indicates their results favour using LLMs, summarising their findings as follows:

> "on average, the 18 LLMs classified 4294 (min 4130; max 4329), 1539 (min 1449; max 1574), and 27 (min 22; max 37) of the titles and abstracts correctly as either included or excluded for the three SRs, respectively".

This statement is misleading because the reviews SR-I and SR-II are highly imbalanced (many irrelevant vs. few relevant studies), meaning rejecting all studies would still achieve high scores for correctness and percentage correctness. For example, gemma:7b in Table 1 scores 96.17% for Accuracy but found *none* of the relevant studies (TP = 0). Similarly, mistral-nemo:12b scored 96.11% for Accuracy, found none of the relevant studies, and identified 3 FPs (i.e., irrelevant studies) as relevant. In contrast, two models that at least identified some of the relevant papers (TP > 0) had lower Accuracy values. This means that if we optimise on the Accuracy metric, we would select worse, or in some cases, completely ineffectual LLMs that failed to detect any relevant primary studies.

In addition, Specificity, which is the proportion of all negatives correctly identified, is also misleading for imbalanced data dominated by negatives. As can be seen in Table 1, the two LLMs that did not classify any of the positives correctly had perfect Specificity values.

### 2.2. The critical importance of lost evidence

In SRs, falsely rejecting relevant studies (FNs) loses evidence, potentially irretrievably. Recall, also referred to as Sensitivity (the proportion of relevant studies identified), and Lost Evidence (1-Recall) are therefore fundamental performance metrics. DC+ Fig. 1 reveals all reviews — even the balanced SR-III — had problematic Lost Evidence scores. Fig. 1 shows that Lost Evidence was problematic across all models, ranging from 14% (best-case in SR-II) to 100% (worst case in SR-I), with 46 out of 54 LLM classifications missing more than 50% of positive papers. No model performed well on all datasets. The only consistency was llama3.2:3b delivered extremely poor predictions on all three data sets, likely because it had the fewest parameters.

Unlike FNs, FPs (irrelevant studies incorrectly included) only waste effort in subsequent screening. The dominant risk to SR validity comes from missing papers that should be included. However, allowing an extremely large number of FPs, in order to minimise FNs, would mean that there was little value in using the LLM classification. This can be modelled as the FN/FP cost ratio, requiring a subjective assessment of the cost of missing evidence versus the cost saving involved in not processing irrelevant studies. The cost ratio will vary by domain and review type (e.g., systematic scoping reviews may tolerate missing studies better than formal SRs), and should be determined by stakeholders based on the specific context and consequences of missed evidence.

Table 1 uses a plausible 10:1 cost ratio, which indicates that Llama3.1:8b provides better classifications than Llama3-Athene:70b, because although it delivers substantially more FPs, it also delivers more TPs and fewer FNs. In addition, Llama3-Athene:70b fails to classify 5 items.

### 2.3. Reporting biased and unsuitable performance metrics

DC+ reported analysis using six performance metrics (Precision, Recall, Specificity, F1, MCC, and PABAK) calculated from the confusion matrices, stating that "multiple evaluation metrics offers a comprehensive perspective on their performance and robustness". However, Precision, Recall, Specificity, and F1 are all biased for imbalanced data. While Recall directly relates to Lost Evidence assessment, the value of the other metrics can all be significantly impacted by imbalanced

**Table 1**
Performance metrics for four of the LLMs used in SR-I, revealing problems with using Accuracy and other non-chance adjusted metrics, and ignoring relative costs.

| Metrics: | Models: | | | |
|---|---|---|---|---|
| | gemma:7b | llama3-Athene:70b | llama3.1:8b | mistral-nemo:12b |
| True Negatives (TNs) | 4324 | 4242 | 4048 | 4326 |
| False Negatives (FNs) | 172 | 125 | 90 | 172 |
| True Positives (TPs) | 0 | 47 | 82 | 0 |
| False Positives (FPs) | 0 | 82 | 281 | 3 |
| Total Articles (N*) | 4496 | 4496 | 4501 | 4501 |
| Evidence Lost | 100% | 73% | 52%* | 100% |
| Accuracy | 96.17%* | 95.40% | 91.80% | 96.11% |
| MCC | NaN | 0.29* | 0.29* | −0.005 |
| Weighted MCC (WMCC)** | NaN | 0.40 | 0.48* | −0.014 |
| Precision | NaN | 0.36* | 0.23 | 0.00 |
| Recall | 0.00 | 0.27 | 0.48* | 0.00 |
| Specificity | 1.00* | 0.98 | 0.94 | 1.00* |
| F1 | NaN | 0.31* | 0.31* | NaN |
| Cost | 1720 | 1332 | 1181* | 1723 |

NB:
* Denote the 'best' LLM by metric, i.e., row-wise.
** Denotes that we used a weight $w = 10$ to calculate WMCC.

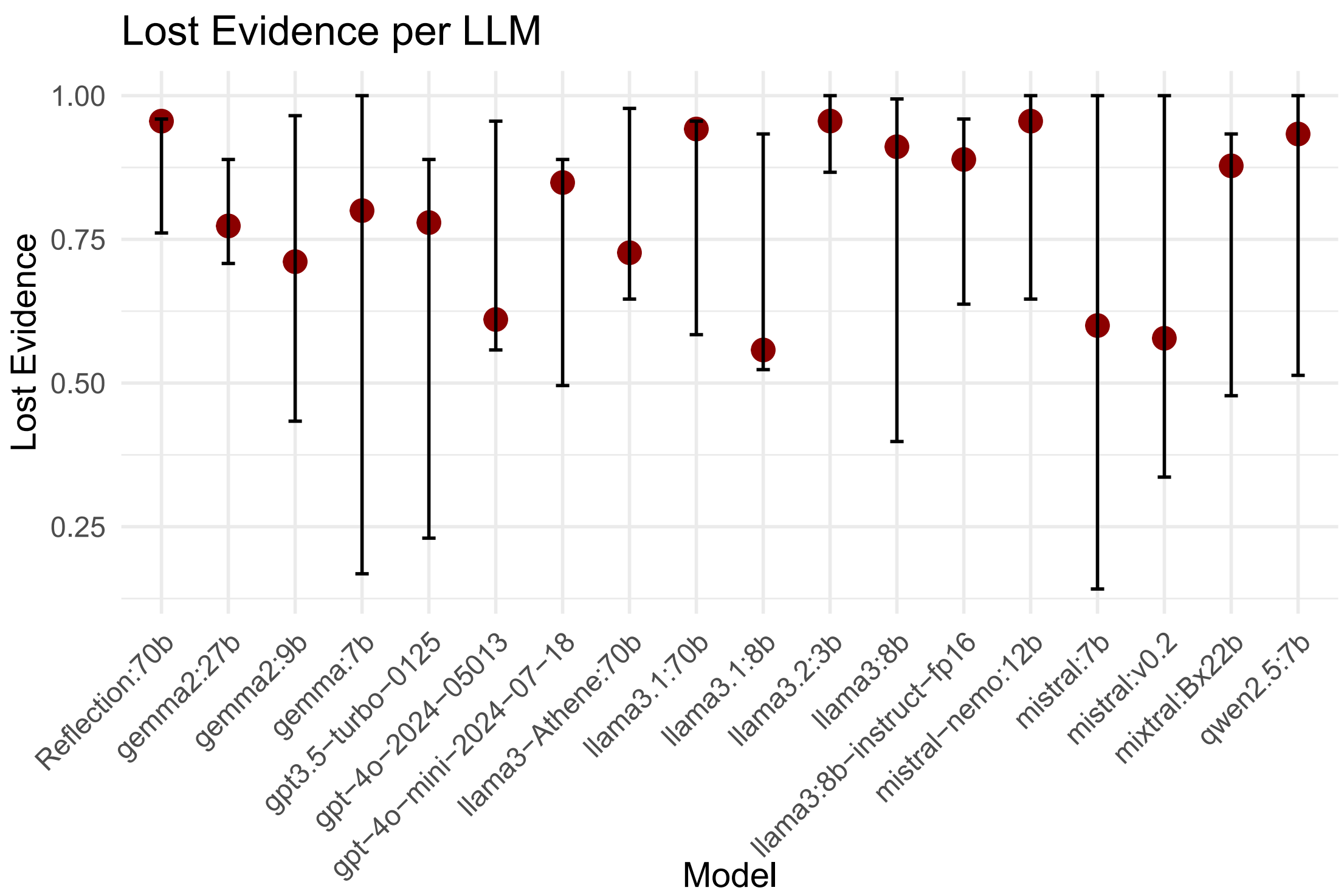

**Fig. 1.** Lost Evidence per Model for three SRs (the median of Lost Evidence is presented as a point and the min/max show the extremes).

data [17]. PABAK, Prevalence Adjusted Bias Adjusted Kappa, [18,19] is equivalent to the centred version of the Accuracy metric, so it is not unbiased in any meaningful way.

Of the metrics DC+ deploy, only the Matthews Correlation Coefficient (MCC) [20] is unbiased because it considers all four elements of the confusion matrix without any bias towards TPs or TNs. MCC is an application of the Pearson correlation coefficient, ranging from −1 to 1, with values near zero indicating chance-level performance. Like any correlation coefficient, MCC remains reasonably robust to imbalanced datasets [20,21]. The problem is that MCC does not address the differential costs of FPs and FNs. However, it is possible to construct an appropriate weighted measure of MCC (WMCC), which we discuss in Section 3.3.

In addition, reporting multiple performance metrics without adequate interpretive framing can be problematic because all these metrics are derived from the same four confusion matrix elements, and the elements of the confusion matrix are not themselves independent. Multiple metrics can be useful if properly contextualised, but the functional correlations between them are often more difficult to understand than the basic confusion matrix elements. This is clear because we only require limited information about the overall classification process in order to generate all four elements. For example, if we know the number of negative papers (N), the number of positive papers (P) as defined by the baseline (gold standard) classification process, together with the number of papers classified as negative by the LLM (n) and the number of those n papers that were TNs, then we can construct the remaining three elements of the confusion matrix because:

$$FP = N - TN \quad (1)$$

and

$$FN = n - TN \quad (2)$$

and

$$TP = P - FN \tag{3}$$

### 2.4. Dropping unclassified papers

DC+ chose to exclude papers that could not be classified from confusion matrices. In real SRs, difficult-to-classify papers typically undergo further screening [22,23]. To align with standard SR practices, LLMs should identify unclassifiable papers as *referred-back* to a human reviewer for further assessment [24]. For the purposes of assessing LLM performance, referred-back papers are a mixture of FPs and TPs (as any referred-back paper will be included in the next screening round, i.e., it will be treated as a positive) and should be classified appropriately in confusion matrices, rather than reducing the total number of classified papers to ignore referred-back papers.

### 2.5. Good evaluation practices

While DC+ illustrates the metric selection challenges discussed above, the paper also adopts two extremely useful good evaluation practices:

**(P1) Reporting full confusion matrices**: DC+ provides the full confusion matrices for each LLM and SR in publicly accessible supplementary materials. Access to the full confusion matrices means that other researchers and meta-analysts can easily construct any performance metric of interest in their own context, in particular, the unbiased MCC metric or the weighted MCC metric (see Section 3.3).

**(P2) Comparing with non-LLM baselines**: Non-LLM baselines complement LLM evaluation when the study goals and study design support them (typically SR updates with prior screening labels, or retrospective benchmarks that include labelled datasets). In particular, non-LLM baselines are often informative when the study aims to assess whether LLMs can support or integrate with existing SR practice. In contrast, for studies whose sole aim is to compare LLMs' performance against each other, or to evaluate different prompting strategies, non-LLM baseline comparisons are unnecessary and P2 can be ignored.

## 3. Current LLM literature screening evaluation practices

To assess whether the performance metric problems in the DC+ paper were representative of current research practice, we also reviewed 28 additional papers on literature screening, including any supplementary materials (see Table 2).

The papers were assembled from three different sources, which are shown in the columns labelled *Origin*. Papers were obtained from two systematic reviews of papers reporting empirical studies of LLM support for literature selection: Kim et al. [15] and Sandner et al. [16]. Kim et al. undertook a meta-analysis of the performance metrics F1, Precision, and Recall/Sensitivity based on 14 relevant papers (including one SE paper [2]) that together reported 33 separate results. Sandner et al. undertook an SR using the performance metrics Recall/Sensitivity and workload reduction. They identified 11 relevant papers (none being SE related) reporting 13 separate results. Five papers were included in both SRs (labelled as Both in the Origin column). In addition, we identified 14 relevant papers (including DC+) from our own informal searches. These papers are labelled A (for Authors) in the Origin column of Table 2. Five of the papers we found were also identified by Kim et al. [15] or Sandner et al. [16]. In addition to [1], we initially identified another five SE-related papers (i.e., [2–6]), bringing the total number of papers assessed to 29.

Table 2 reports details about the papers included in this review:

1. The column labelled *Performance Metrics*, identifies the metrics reported in each paper, excluding simple confusion metric counts. The performance metrics referred to by shortened labels are Acc (Accuracy or Correctness), BalAcc (Balanced Accuracy), Rec (Recall or Sensitivity), Spec (specificity), Prec (Precision), NegPred (Negative Prediction), RecInc (Sensitivity Included), RecExc (Sensitivity Excluded, Pos (Positive), RecAT% (Recall at % checked from an ordered list), Neg (Negative), Null (Number of null or otherwise invalid outcomes), PropMissed (Proportion Missed). The term *WS* was used to refer to some form of work saved metric irrespective of the specific term used by the authors. In addition, Akinseloyin et al. [25] investigated prioritising papers in terms of relevance, rather than classifying them. They used metrics appropriate to that task, but did not fully define them. Their Recall statistics and Work saved statistics were calculated relative to the percentage of prioritised papers evaluated.
2. The seven papers that reported confusion metric counts or percentages (in the paper itself or in supplementary material from which counts can be reconstructed) are identified in the column labelled *CM*.
3. The column labelled *Type* identifies whether the paper was published in a journal (J), conference proceedings (C), was only available as an Abstract (A), was a letter to the editor (L), or was grey literature (GL).
4. The column labelled *Other Baselines* identifies the papers that compared LLM performance with other types of machine learning algorithms, such as logistic regression.

Based on the Performance Metrics column, Fig. 2 reports a summary of the performance metric usage.

### 3.1. Review results

The 29 papers (28 other papers and DC+) reveal significant methodological gaps in evaluating Gen-AI tools for systematic review screening, with an underutilisation of appropriate unbiased performance metrics such as MCC (see Fig. 2) and limited approaches to cost–benefit considerations:

1. **Limited MCC adoption**: Only 3 papers (10% of the sample) employed MCC, including DC+ and the two papers by Syriani et al. [5,6], despite MCC's advantages for handling imbalanced datasets common in systematic review screening. However, Syriani et al. rescaled MCC from its standard $[-1, 1]$ range to $[0, 1]$, which may complicate cross-study comparisons. Failure to use MCC is a missed opportunity for robust performance evaluation, particularly given the inherent class imbalance in screening tasks.
2. **Insufficient confusion matrix reporting**: After searching not only the papers but also any reported supplementary material, we found only 7 papers reported complete confusion matrices, though 4 additional papers provided sufficient information (in terms of total positives, total negatives, sensitivity, and specificity) to enable reconstruction. This deficiency hampers meta-analysis by preventing the construction of MCC, which is well-suited for meta-analysis (see [44]) and prevents readers from computing alternative metrics or conducting independent cost–benefit assessments tailored to their specific screening contexts. It should also be noted that reporting the proportions or percentages in each confusion metric element without reporting the total number of items also limits analysis, because without the raw counts, although MCC and WMCC can be calculated, a full cost–benefit analysis is impossible. As a further example of its importance, we also note that two of the 24 secondary studies in the SESR-Eval dataset [2] contain only included papers (zero excluded), meaning TN = 0 and FP = 0 by construction. For these studies, Precision is trivially 1.0 whenever any paper is correctly included, which inflates F1 and biases aggregate performance statistics — further illustrating why reporting complete confusion matrices is essential, as it makes such degenerate cases immediately visible.

**Table 2**
Summary of the screening papers.

| ID | Performance metrics | CM | Origin | Type | Other baselines |
|---|---|---|---|---|---|
| Akinseloyin-2024 [25] | L-Rel; MAP; RecAT%; WS | No | Sandner | J | No |
| Attri-2024 [26] | Acc; %Pos; Rec; Spec | No | Kim | A | No |
| Cai-2023 [27] | F1; Prec; Rec; WS | No | Both | J | No |
| Cao-2024 [28] | Acc; Rec; Spec | No | Sandner; A | GL | No |
| Castillo-2023 [29] | Acc; F1; NegPred; Nulls; Prec; Rec; Spec | Yes | A | C | No |
| Datta-2024 [30] | Acc; F1; Prec; Rec | No | Kim | A | No |
| DC+ [7] | Corr; F1; MCC; PABAK; Prec; Rec; Spec | Yes | A | A | Yes |
| Dennstadt-2024 [8] | Acc; F1; Prec; Sens; Spec | Yes | Kim | J | Yes |
| Du-2024 [31] | Acc; F1; Prec; Rec | No | Kim | J | Yes |
| Felizado-2024 [4] | Acc | Yes | A | J | No |
| Gargari-2024 [32] | Acc; F1; Rec; Spec | No | Sandner | L | No |
| Guo-2024 [33] | Acc; F1; Kappa; PABAK; RecExc; RecInc | Yes | Both;A | J | No |
| Huotala-2024 [1] | %Exc; %Inc; Prec; Rec | No | Kim; A | J | No |
| Huotala-2025 [2] | Acc; F1; Prec; Rec | Yes | A | C | No |
| Issaiy-2024 [34] | BalAcc; Jaccard; Kappa; NegPred; Pos&Neg Likelihood; Prec; PropMissed; Rec; Spec; WS | No | Both | J | No |
| Kaur-2024 [35] | Acc; Rec; Spec | No | Kim | A | No |
| Khraisha-2024 [9] | Acc; Kappa; PABAK; Rec; Spec; Weighted Kappa | No | Both;A | J | No |
| Li-2024 [36] | Acc; Rec; Spec | No | Sandner | J | No |
| Lin-2023 [37] | Acc; F1; Prec; Rec; ROC; Spec | No | Kim | J | Yes |
| Rai-2024 [38] | Acc; Prec | No | Kim | A | Yes |
| Robinson-2024 [39] | Acc; Prec; Rec | No | A | GL | Yes |
| Royer-2023 [40] | Rec; Spec | No | Kim | A | Yes |
| Spillias-2024 [41] | Kappa | Yes | Sandner | J | No |
| Syriani-2023 [5] | BalAcc; F2; Fleiss' Kappa; MCC; NegPred; Prec; Rec; Spec | No | A | GL | Yes |
| Syriani-2024 [6] | BalAcc; MCC; NegPred; Prec; Rec; Spec | No | A | J | Yes |
| Thode-2025 [3] | Prec; Rec | No | A | J | No |
| Tran-2023 [42] | Rec; Spec; WS | No | Sandner | GL | No |
| Wang-2024 [13] | BalAcc; F3; Prec; Rec; Success Rate; WS | No | Both;A | C | No |
| Wilkins-2023 [43] | Acc; Kappa; Weighted Kappa; Weighted Rec; Weighted Spec | No | A | GL | No |

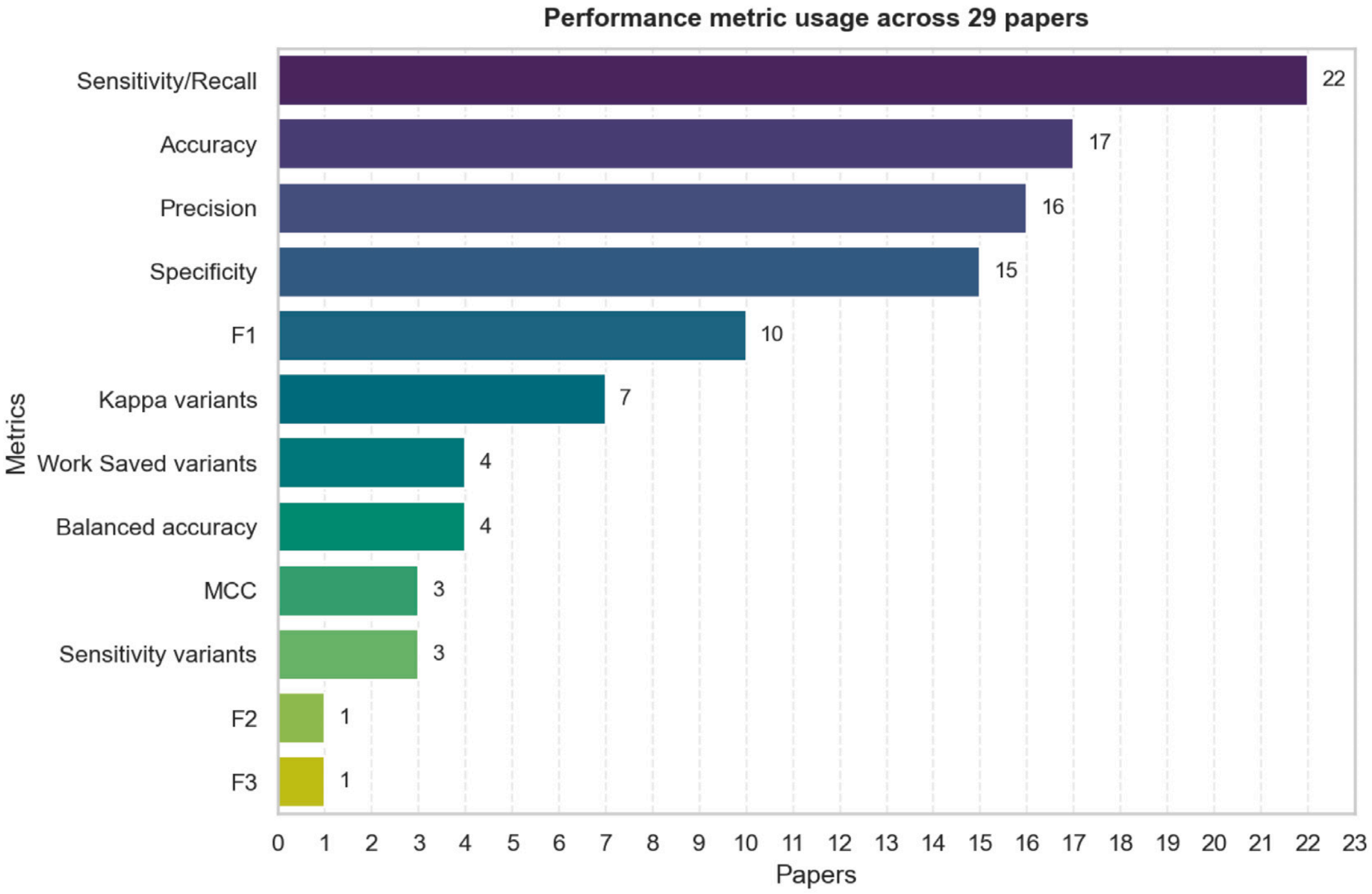

**Fig. 2.** Distribution of evaluation metrics used across 29 papers analysing Gen-AI tools for systematic review screening.

3. **Sensitivity/Recall and Accuracy as dominant metrics**:

- 22 papers used standard Recall/Sensitivity, while the additional 3 papers reported some form of recall-related metric, including weighted variants and partial metrics such as RecAT%, RecInc/RecExc, or Weighted Recall. While this prevalence is encouraging and reflects the field's appropriate concern with minimising FNs, the use of recall often occurred without complementary chance-anchored metrics (such as MCC) to assess the trade-offs with overall efficiency, and none of the five papers reporting workload savings incorporated FN costs into their calculations.

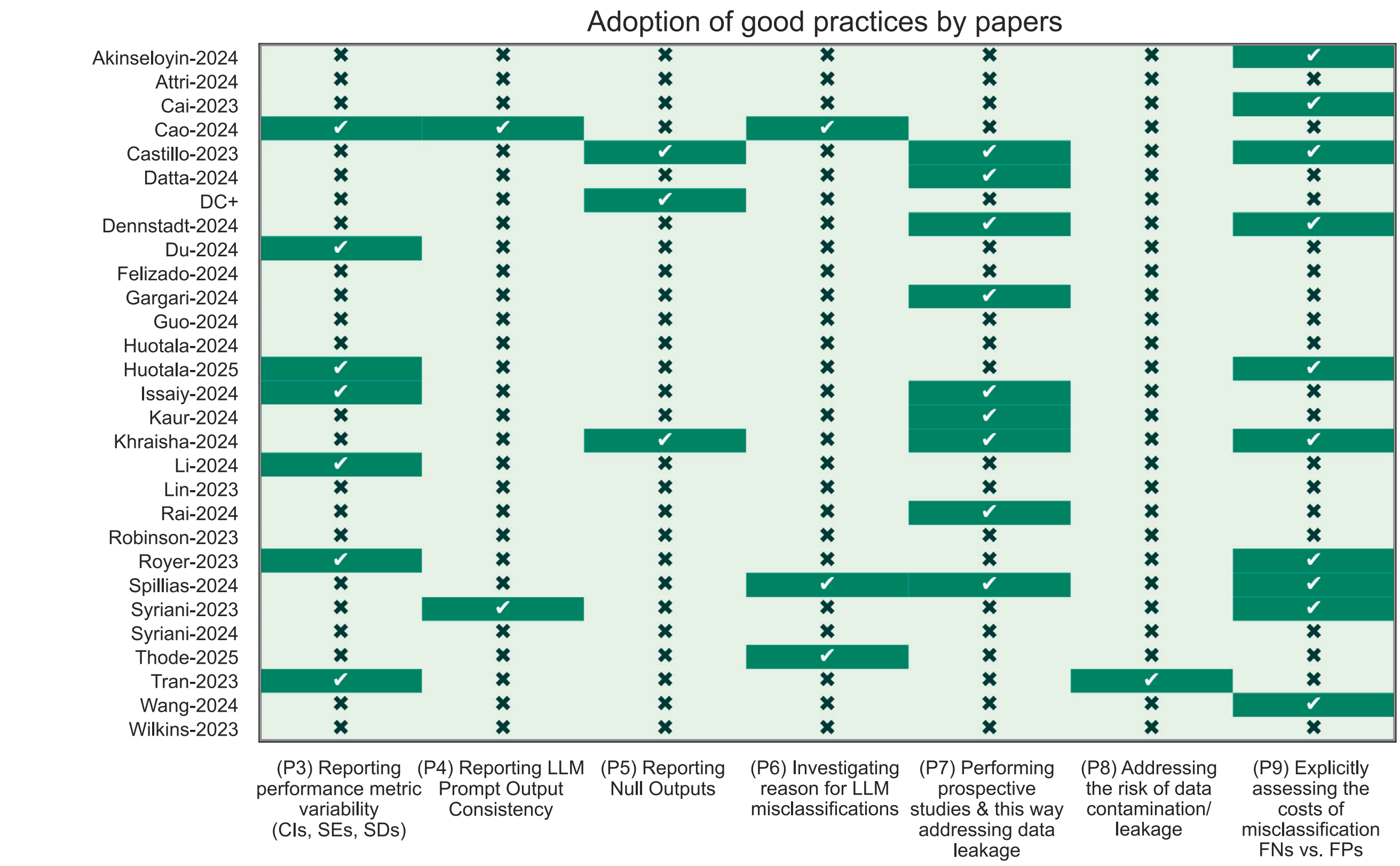

**Fig. 3.** Adoption of good practices.

- The widespread use of accuracy (17 papers) alongside high sensitivity reporting suggests many researchers may not fully appreciate accuracy's limitations for imbalanced datasets.
- The very modest adoption of balanced accuracy (only 4 papers, 14%) and extremely low MCC usage indicates insufficient awareness of metrics specifically designed for class imbalance scenarios or the need for chance-anchoring.
- Seven papers used Kappa variants (some employing multiple versions), which is appropriate when there is no well-defined baseline. However, it is inappropriate to use standard performance metrics and kappa variants on the same confusion matrix because (with the exception of MCC) standard confusion matrix metrics assume that a gold standard exists, while kappa assumes that no gold standard exists. Kappa is appropriate when comparing two potentially fallible classifications (e.g., classifications made by a single human and/or one or more LLMs). In addition, the PABAK variant does not seem to be a useful metric in any circumstances.

4. **Failure to address costs as well as benefits:** None of the five papers that explicitly claimed to measure workload savings, nor the Sandner SR [16] that explicitly investigated workload savings, considered the cost of False Negatives when specifying their WS metric.

### 3.2. Additional good practices

In addition to two good practices (P1 and P2) already found in DC+ [7] and reported in Section 2.5, we also observed several other good (evaluation) practices which are shown in Fig. 3:

**(P3) Reporting performance metric variability**: When conducting a new SR, current SR standards (both medical [45] and SE [22]) mandate that any GenAI tool is validated and the validation is reported. Validation would need to be based on a sample of the papers to be screened. Researchers can then compare the performance of different LLMs with that obtained from human researchers. Confidence limits attached to the performance metrics are necessary to identify which LLMs (singly, or in combination with other LLMs or a single human, or also any other more deterministic tools) are most likely both to meet agreed performance levels and to introduce work load reductions of sufficient magnitude to balance potential loss of evidence.[2] Seven papers reported performance metric confidence intervals, standard errors, or standard deviations, and five of the papers were undertaking studies based on samples. However, there was little consistency in the methods used to calculate confidence intervals. Two papers used properties of the binomial distribution, one used bootstrapping, and the others (apart from the paper by [2], which reported standard errors obtained from logarithmic regression) did not specify any particular method.

**(P4) Reporting LLM prompt output consistency (optional) & (P5) Reporting null outcomes**: Two papers considered the internal consistency of LLMs, and three other papers reported the rate of null outcomes. While detailed consistency analysis (varying temperature, top-k, etc.) may warrant separate investigation, basic reporting of output consistency is valuable for reproducibility. Null outcomes affect performance assessment regardless of whether they originate from the LLM itself or from infrastructure issues (e.g., network problems, API downtimes); both their rate and suspected cause should be reported, as they directly affect confusion matrix computation.

[2] In the context of retrospective studies based on existing SRs, the classification data is available for all the abstracts i.e., the population of abstracts for a specific SR have all been classified, so the standard error of any performance metric obtained from testing LLMs on the full data set is zero and calculating confidence intervals invalid within that SR; CIs remain valid when the benchmark uses subsampling or pools across SRs (see Section 3.5 and R5).

**(P6) Investigating the reason for LLM misclassifications (optional, but recommended for iterative studies)**: Three papers assessed the reason for LLM classifications and found systematic problems with their prompts. This practice is particularly valuable for iterative studies where findings can inform prompt refinement. We acknowledge that it represents a shift from pure evaluation towards LLM improvement, and is therefore optional.

**(P7) Performing prospective studies & this way addressing data leakage**: Nine studies were prospective studies (as opposed to retrospective studies, which re-analyse previously published studies). Generally, prospective studies are preferable because there is less chance that researchers have oriented the study goals to suit the available data. In addition, prospective studies do not suffer from the risk of data leakage, also known as data contamination [24].

**(P8) Addressing the risk of data contamination/leakage**: In the context of LLM testing, there is always a risk that publicly available data used to train an LLM could be part of the data used to test LLM performance [24]. This is referred to as data leakage or data contamination. Only Tran et al. [42] reported that their retrospective study used data published after the LLM they studied was released. Dennstadt et al. [8], who used one prospective study and 10 benchmark studies, mentioned the issue and pointed out that "To objectively assess how well an LLM-based solution can evaluate scientific publications for new research questions, large cultivated and independent prospective data sets on many different topics would be needed, which will be very challenging to create". However, in general, there was little appreciation of the issue. None of the other seven papers that used data from public benchmarks mentioned the issue, and three papers that used retrospective studies that were not obtained from public benchmarks suggested that their data sets could be used as benchmarks.

**(P9) Explicitly assessing and documenting the differential costs of FNs vs. FPs**: None of the papers reporting work-saving metrics incorporated FN into their metric. They reported only the savings due to not requiring TNs to be processed and costs due to the additional processing of FPs. Failing to consider the risks posed by FNs is unlikely to lead to balanced assessments of the performance of different GenAI tools. In contrast, ten papers explicitly considered the differential costs of FNs vs. FPs, but with substantial variations in their approaches:

- Khraisha-2024 [9] assigned FNs a weight 30 times greater than FPs, representing the most aggressive cost difference.
- Wang-2024 [13] mandated a minimum 95% recall threshold, prioritising sensitivity over other metrics.
- Syriani et al. [5] employed F2 scores (weighting recall twice as heavily as precision), although they did not use F2 as a performance metric in their subsequent paper [6]. However, in both papers, their prompts requested Gen-AI systems to be lenient towards inclusions.
- Huotala et al. [2] suggest setting a target of 95% for recall, with precision of approximately 50%.
- The other six papers mentioned either the critical importance of Recall or the danger of missing evidence, but did not suggest any specific evaluation practices to address the issue.

### 3.3. Addressing performance metrics limitations

In this paper, we have identified the problems that arise when confusion matrices are strongly imbalanced, performance metrics do not consider all elements of the confusion matrix, and do not have a meaningful zero that corresponds to a classifier that is not performing better than chance. We have also noted that the Matthews Correlation Coefficient addresses these issues. In addition, it also allows formal statistical tests to indicate whether or not a given MCC value is better than a random classifier.

However, we have also criticised performance metrics that do not address the cost asymmetry of FNs vs. FPs and, MCC, as specified in Eq. (15), clearly does not address this issue. Thus, to address cost asymmetry, we propose using a **Weighted Matthews Correlation Coefficient (WMCC)** that builds upon ordinary MCC.

The idea behind WMCC is that it preserves MCC's chance-anchored,[3] imbalance-robust correlation meaning and directly addresses the cost asymmetry, although at the cost of losing the opportunity to perform the customary MCC statistical tests of significance. The general WMCC formula is presented in Eq. (4):

$$WMCC = \frac{(TP_w * TN_w - FP_w * FN_w)}{\sqrt{(TP_w + FP_w) * (TP_w + FN_w) * (TN_w + FP_w) * (TN_w + FN_w)}} \quad (4)$$

Constructing a class-weighted version of MCC, i.e., WMCC, we assign weight $w$ (e.g., $w = 10$) to each positive example, i.e., TP and FN, and weight 1 to each negative example, i.e., TN and FP (when positives are $w$-times more consequential than negatives), compute the weighted confusion matrix counts, and plug them into the standard MCC formula.

Hence, weighted counts are as in Eq. (5):

$$TP_w = w \cdot TP, \quad FN_w = w \cdot FN, \quad TN_w = 1 \cdot TN, \quad FP_w = 1 \cdot FP \quad (5)$$

and WMCC can be simplified to the form presented in Eq. (6):

$$WMCC = \frac{(w * TP * TN - FP * w * FN)}{\sqrt{(w * TP + FP) * (w * TP + w * FN) * (TN + FP) * (TN + w * FN)}} \quad (6)$$

Selecting the weight $w$ requires consideration of the specific SR context. We recommend: (1) stakeholder consultation to determine domain-specific consequences of missed evidence versus wasted screening effort; (2) sensitivity analysis to assess how different plausible values of $w$ affect model rankings; and (3) a default of $w = 10$ as a reasonable starting point for many SR contexts, reflecting that missing a relevant study typically has greater consequences than including an irrelevant one for further review. The chosen weight should be documented and justified in the study protocol. For deployment studies with access to domain stakeholders, approach (1) is the primary method. For benchmarking studies without domain-specific stakeholders, sensitivity analysis (2) is the preferred approach as it reveals whether model rankings are robust to cost-ratio assumptions; the default $w = 10$ (3) can additionally be reported as a cross-study reference point. Empirical support for $w = 10$ as a conservative default comes from the SE-specific sensitivity analyses in Section 3.4: the 12 SESR-Eval crossovers (median $w \approx 2.7$, max $w \approx 6.4$) — and the illustrative Syriani RL4SE crossover ($w \approx 6$) — all occur well below $w = 10$.

A simple, working example (for two LLMs, `llama3-Athene:70b` and `llama3.1:8b`, from Table 1) of how to calculate WMCC under class imbalance, with asymmetric costs reflected by a weight of $w = 10$, is presented below.

[3] Performance metrics are considered chance-anchored if a defined and stable point corresponds to random performance, so for a correlation metric this is zero and for AUC this is 0.5.

**Table 3**
SE-specific worked example: reanalysis of ChatGPT-4 screening results from Felizardo et al. [4] (ESEM'24) on two SE SLRs (using Felizardo et al.'s original dataset names). Confusion-matrix counts are taken verbatim from [4]; MCC and WMCC ($w = 10$) are computed by us. WMCC weights each positive example (TP, FN) ten times more than each negative example (TN, FP).

| SLR | Threshold | Confusion matrix | | | | N | Acc | Recall | Lost Ev. | MCC | WMCC |
|---|---|---|---|---|---|---|---|---|---|---|---|
| | | TP | FP | FN | TN | | | | | | |
| SLR1 | Likert ≥ 4 | 50 | 35 | 14 | 35 | 134 | 63.4% | 0.781 | 21.9% | 0.292 | 0.195 |
| SLR1 | Likert ≥ 5 | 48 | 17 | 16 | 53 | 134 | 75.3% | 0.750 | 25.0% | 0.507 | 0.330 |
| SLR2 | Likert ≥ 4 | 128 | 68 | 20 | 232 | 448 | 80.3% | 0.865 | 13.5% | 0.605 | 0.557 |
| SLR2 | Likert ≥ 5 | 113 | 27 | 35 | 273 | 448 | 86.1% | 0.764 | 23.6% | 0.683 | 0.529 |

Raw counts for `llama3-Athene:70b` LLM from Table 1: $TP = 47$, $FN = 125$, $TN = 4242$, $FP = 82$. Assuming $w = 10$, we may calculate WMCC for this LLM:

$$WMCC^{w=10}_{llama3-Athene:70b} = \frac{(10*47*4242-82*10*125)}{\sqrt{(10*47+82)*(10*47+10*125)*(4242+82)*(4242+10*125)}} = \frac{1891240}{4748341} = 0.398 \quad (7)$$

Raw counts for `llama3.1:8b` LLM from Table 1: $TP = 82$, $FN = 90$, $TN = 4048$, $FP = 281$. Assuming $w = 10$, we may calculate WMCC for this LLM:

$$WMCC^{w=10}_{llama3.1:8b} = \frac{(10*82*4048-281*10*90)}{\sqrt{(10*82+281)*(10*82+10*90)*(4048+281)*(4048+10*90)}} = \frac{3066460}{6368931} = 0.481 \quad (8)$$

This example shows the impact of weighting MCC. WMCC allows us to distinguish between `llama3-Athene:70b` and `llama3.1:8b`, which both had the same MCC value (see Table 1), indicating that `llama3.1:8b` is a better classifier because $WMCC^{w=10}_{llama3.1:8b} > WMCC^{w=10}_{llama3-Athene:70b}$ due to `llama3.1:8b` having fewer FNs than `llama3-Athene:70b`, although it has substantially more FPs.

### 3.4. SE-specific worked examples

The preceding analysis uses confusion matrices from the biomedical DC+ study. To ground the same analysis in software engineering, we reuse the only SE paper in our review that reports complete confusion-matrix counts directly in the paper: Felizardo et al. [4] (ESEM'24), who evaluated ChatGPT-4 on two SE SRs at two classification thresholds (Likert ≥5 as the primary threshold and Likert ≥4 as a "conservative" threshold that favours inclusion). Table 3 reports the raw counts from [4] together with Accuracy, Recall, Lost Evidence, MCC, and WMCC ($w = 10$) that we computed from those counts.

Four observations emerge from this SE-specific reanalysis:

1. *Accuracy and MCC can prefer the wrong operating point.* For SLR2, Accuracy favours the Likert ≥5 threshold (86.1% vs. 80.3%), and MCC agrees (0.683 vs. 0.605). Yet Recall (0.865 vs. 0.764), Lost Evidence (13.5% vs. 23.6%), and WMCC (0.557 vs. 0.529) all favour the more inclusive Likert ≥4 threshold. A practitioner who optimised on Accuracy or even on MCC would select the threshold that loses almost twice as much evidence (23.6% vs. 13.5%) — MCC corrects for chance but, because it treats FNs and FPs symmetrically, it still prefers the more specific threshold, whereas WMCC's 10:1 FN weight correctly rewards the higher-Recall operating point.
2. *WMCC can tolerate a small increase in Lost Evidence when workload reduction is large.* For SLR1, the Likert ≥5 threshold has two more FNs (16 vs. 14) but eighteen fewer FPs (17 vs. 35) and eighteen more TNs (53 vs. 35) than Likert ≥4. Under a simple 10:1 weighted-cost model (FP + 10·FN), Likert ≥5 is marginally more costly (177 vs. 175), yet WMCC strongly prefers Likert ≥5 (0.330 vs. 0.195) because, unlike the simple weighted cost, chance-anchored metrics credit the large increase in correctly excluded abstracts (TN). WMCC therefore rewards screening-workload reductions that are large enough to compensate for a small increase in Lost Evidence — the cost asymmetry operates through the confusion matrix as a whole, not through FN counts alone.
3. *MCC is chance-anchored; Accuracy is not.* All four configurations show Accuracy between 63% and 86%, yet MCC ranges from 0.29 to 0.68 — the chance-anchored metric reveals that the classifier's improvement over random guessing is more modest than Accuracy suggests.
4. *WMCC penalises FNs more severely than MCC.* In every row, WMCC($w = 10$) < MCC, because the 10:1 FN penalty pulls the cost-sensitive score below the symmetric one — demonstrating that WMCC is doing what it advertises.

A nuance is that whilst Felizardo et al. report that only 2 of SLR1's 16 FNs and 4 of SLR2's 35 FNs ultimately remained as lost evidence after the full-text screening stage [4], this is simply a property of the retrospective SR pipeline (later stages can recover some FNs), not of the LLM-tool itself, and motivates the discussion of extending recommendations to later SR stages in Section 4.3.

We also note that none of the SE studies reanalysed in this section (Felizardo et al. Syriani et al. Huotala et al.) address data contamination risks (P8/R8) since the SE papers being screened may have been in the LLMs' training data. This differs from Tran et al. [42] who intentionally selected SRs published after the LLM's training cutoff. We address this concern empirically below via a cross-domain reanalysis of DC+ (see *Contamination-risk sensitivity analysis*).

Beyond this worked example, the remaining SE papers in our review further illustrate the challenges and good practices we recommend. Note we include forward pointers to our researcher and practitioner recommendations detailed in Section 4.2 to assist cross-referencing for the reader.

- **Huotala et al. [1] (EASE'24):** found their zero-shot GPT-3.5 and GPT-4 missed 35%–50% of included papers when reproducing a prior SE SR screening — a direct SE-specific Lost Evidence illustration. Only Recall and Precision were reported so no confusion matrix, MCC metrics or cost-sensitive analysis.
- **Huotala et al. [2] (ESEM'25, SESR-Eval):** This was the largest SE-specific benchmark (9 LLMs × 24 secondary studies, 34,528 primary studies). Accuracy ranged 0.34–0.85 and F1 ranged 0.07–0.92 across secondary studies. Two of the 24 secondary studies, [53,64] from their reference numbering, contain only included papers (I/E Ratio = 100:0 in their Table VI and Huotala et al. themselves note that [53] "contains only included studies"). This creates a degenerate case since trivially Precision = TP/(TP + 0) = 1.00 for any LLM that includes at least one paper, and the $(TN+FP)$ factor in the MCC denominator (Eq. (15)) is zero, making MCC mathematically undefined. Including these two studies in per-study averages inflates aggregate Precision by + 0.05 and F1 by + 0.03 relative to the remaining 22 studies (computed from their Table IX) which is exactly the kind of artefact that complete confusion-matrix reporting (R4) makes visible. No MCC or cost-sensitive metrics were reported. We reanalyse the SESR-Eval data below.

- **Thode et al.** [3] **(IST):** Reports only Recall and Precision for three LLMs, three prompt templates, and three SE datasets. A two-LLM ensemble reached 98%–99% recall at 27% precision, however, no confusion matrix, no MCC, no cost-sensitive analysis, and no non-LLM baselines were reported. Data contamination risk was acknowledged but not mitigated, hence this is an SE example where our recommendations could have strengthened the evaluation.
- **Syriani et al.** [5,6]**:** These are the only SE papers in our review reporting MCC and comparing ChatGPT against non-LLM baselines (LR, RF, CNB, SVC via 5-fold CV with grid search), illustrating the SE operationalisation of P2/R9 when training data is available. However, they rescaled MCC from $[-1, 1]$ to $[0, 1]$ using $\text{MCC}_{[0,1]} = 0.5 + \text{MCC}_{[-1,1]}/2$ [6], which inflates apparent performance and complicates cross-study comparisons. We reanalyse their data in Table 4.

**Syriani et al. reanalysis.** Syriani et al. [6] report Recall, Specificity, and the number of included/excluded papers for each of five SE datasets (see their Table 1). For three datasets (UpdateCollabMDE, MobileMDE, MPM4CPS) we verified the confusion matrices directly against Syriani et al.'s public replication package[4]; the counts match exactly. For the remaining two datasets (RL4SE, DSMLCompo) the replication package contains results from a different experimental run, so we approximated the confusion matrices via $TP = \text{round}(\text{Recall} \times P)$ and $TN = \text{round}(\text{Specificity} \times N)$. Because these reported metrics are rounded to three decimal places, the reconstructed TP and TN counts may each differ from the true values by $\pm 1$, propagating to at most $\pm 0.007$ in MCC and $\pm 0.012$ in WMCC — well within the gap that separates classifiers in every comparison below. For the non-LLM baselines, Syriani et al. trained classifiers with 5-fold cross-validation repeated 10 times, so their reported Recall and Specificity are averages across folds and repeats. The confusion matrices we present for baselines are therefore synthetic (derived from averaged metrics) rather than observed from a single evaluation pass; we retain them because they are the only available basis for computing standard MCC and WMCC, but readers should note this caveat. Table 4 shows the results for ChatGPT's best prompt per dataset alongside the best non-LLM baseline, both selected on rescaled MCC as in [6] Table 5.

Three observations emerge from this reanalysis:

1. *Rescaling inflates apparent performance.* Syriani et al.'s rescaled $\text{MCC}_{[0,1]}$ for ChatGPT ranged 0.638–0.767 across the five datasets, suggesting moderate-to-good classification. Standard $\text{MCC}_{[-1,1]}$ ranges only 0.276–0.534 — performance modestly above chance. A naïve cross-study comparison of Syriani's rescaled values with, e.g., Felizardo's standard MCC (Table 3) would be seriously misleading, which is why R1 recommends reporting MCC unrescaled in $[-1, 1]$.
2. *MCC and WMCC can disagree on the winner.* For RL4SE, standard MCC favours LR over ChatGPT (0.347 vs. 0.298), but WMCC favours ChatGPT (0.511 vs. 0.485) because ChatGPT's higher Recall (0.821 vs. 0.599) is rewarded by the 10:1 FN weight. A practitioner who chose the classifier using only MCC would select the one that loses 40% of relevant studies instead of 18% — precisely the scenario that cost-sensitive evaluation (R2) is designed to prevent. To verify that this ranking flip is not an artefact of the specific $w = 10$ choice, we computed WMCC for both classifiers across $w \in \{1, 2, \ldots, 20\}$: the crossover occurs at $w \approx 6$, meaning ChatGPT is the preferred classifier for any integer $w \geq 7$. The flip is therefore robust for any cost ratio that values missed evidence at least seven times more than wasted screening effort — a threshold most SR contexts would meet. (Because the RL4SE confusion matrices are approximated, we additionally confirmed that the flip persists across all $\text{TP} \pm 1$ and $\text{TN} \pm 1$ combinations for both classifiers at $w = 10$, that is 81/81 scenarios; see our replication-package script for details.) As the LR baseline confusion matrix is synthetic (derived from CV-averaged metrics), the RL4SE flip is best read as an illustrative single-pair example; the SESR-Eval reanalysis below provides the primary empirical demonstration using real per-article decisions.
3. *Non-LLM baselines contextualise LLM performance.* ChatGPT outperforms the best baseline on both MCC and WMCC in four of five datasets, but LR wins on MCC for RL4SE and all baselines show substantial Lost Evidence (25%–67%). This illustrates R9, that when training data is available, baselines provide an informative reference point, but neither ChatGPT nor the baselines achieve levels of Lost Evidence that would support confident deployment.

**SESR-Eval reanalysis.** To test whether the patterns observed with Felizardo's four configurations and Syriani's five datasets generalise, we computed MCC and WMCC($w = 10$) for all 216 LLM×study cells (9 LLMs × 24 secondary studies) from the SESR-Eval replication package [2].[5] MCC is computable for 183 of the 216 cells; the remaining 33 are undefined, predominantly for Ministral 8B, which classified virtually every article as "include" (Recall = 1.00 but Accuracy as low as 1.2%). Table 5 illustrates the most striking study: a large SE secondary study (9695 articles, 515 included, 9180 excluded) in which Accuracy, MCC, and WMCC each select a *different* LLM as the best classifier.

Three observations emerge from this large-scale reanalysis:

1. *Accuracy misleads at scale.* Accuracy and MCC disagree on the best LLM in 11 of 22 evaluable secondary studies (50%; two studies with only included papers are excluded because MCC is undefined as explained above). In the example of Table 5, the Accuracy-best model (Claude 3.7 Sonnet, 94.4%) loses 63.3% of the relevant evidence; even the MCC-best model (GPT-4.1 mini) still loses 43.9%. Only WMCC selects the model (GPT-4o) that retains 94.2% of the evidence — at the cost of additional FPs that increase human screening workload but do not compromise SR validity. The most extreme Accuracy-misleads instance occurs in a different study (3703 articles, 45 included): GPT-4.1 mini achieves 97.5% Accuracy but loses 80% of the relevant evidence (MCC = 0.154).
2. *The MCC $\neq$ WMCC ranking flip is pervasive, not isolated.* The Syriani RL4SE example (Table 4) showed a single MCC → WMCC flip between two classifiers on one dataset. SESR-Eval reveals this is not an isolated case: MCC and WMCC disagree on the best LLM in 12 of 22 evaluable studies (55%), replicating the ranking-flip pattern at 20× the scale across diverse SE domains. In every disagreement, WMCC favours the higher-Recall, lower Lost Evidence model, precisely the behaviour the 10:1 FN weight is designed to produce. In the most extreme case, the MCC-best model (GPT-4.1 mini, MCC = 0.481, Recall = 0.561) loses

[4] Available at https://doi.org/10.5281/zenodo.10257742. The replication package also reveals that ChatGPT initially produced API errors (labelled "unknown") for 15, 5, and 4 articles in these three datasets respectively. These were all retried and successfully resolved into INCLUDE/EXCLUDE decisions so the confusion matrices in Table 4 reflect the resolved outputs. The initial error rates (1.7%, 1.7%, 2.0%) are an SE-specific instance of the null-output issue addressed by R6.

[5] The R script reproducing this analysis is included in our replication package.

**Table 4**
Reanalysis of Syriani et al. [6]: ChatGPT (API model GPT-3.5-turbo-0613; best prompt per dataset) vs. best non-LLM baseline on five SE datasets. ChatGPT confusion matrices for UpdateCollabMDE, MobileMDE, and MPM4CPS are taken directly from the replication package; those for RL4SE and DSMLCompo are approximated from reported Recall, Specificity, and known class counts (shown in parentheses as positives+/negatives-), with counts accurate to ±1. Baseline confusion matrices are synthetic, derived from cross-validation-averaged metrics. MCC is on the standard $[-1, 1]$ scale; WMCC uses $w = 10$. Compare with Syriani et al.'s rescaled $\mathrm{MCC}_{[0,1]}$: ChatGPT ranged 0.638–0.767, baselines 0.584–0.734.

| Dataset | Classifier | Confusion matrix | | | | Lost Ev. | MCC | WMCC |
|---|---|---|---|---|---|---|---|---|
| | | TP | FP | FN | TN | | | |
| RL4SE | ChatGPT | 77 | 310 | 17 | 685 | 18.1% | 0.298 | 0.511 |
| (94+/995-) | LR | 56 | 130 | 38 | 865 | 40.4% | 0.347 | 0.485 |
| DSMLCompo | ChatGPT | 114 | 572 | 36 | 1961 | 24.0% | 0.281 | 0.522 |
| (150+/2533-) | CNB | 112 | 790 | 38 | 1743 | 25.3% | 0.211 | 0.421 |
| UpdateCollab. | ChatGPT | 51 | 286 | 6 | 532 | 10.5% | 0.276 | 0.542 |
| (57+/818-) | SVC | 27 | 123 | 30 | 695 | 52.6% | 0.212 | 0.353 |
| MobileMDE | ChatGPT | 47 | 50 | 8 | 187 | 14.5% | 0.534 | 0.624 |
| (55+/237-) | SVC | 40 | 45 | 15 | 192 | 27.3% | 0.463 | 0.497 |
| MPM4CPS | ChatGPT | 79 | 31 | 28 | 67 | 26.2% | 0.423 | 0.256 |
| (107+/98-) | RF | 35 | 18 | 72 | 80 | 67.3% | 0.164 | 0.086 |

**Table 5**
SESR-Eval reanalysis: Accuracy, MCC, and WMCC ($w = 10$) each select a different "best" LLM on the same 9695-article SE secondary study (515 included, 9180 excluded). Bold values mark the metric on which each model ranks first. The `gpt-4.1-nano` row shows the model a pure Recall-maximiser would select. Ministral 8B is included for contrast: it achieves 100% Recall by classifying every article as "include", producing undefined MCC. Confusion matrices are computed from the raw per-article decisions in the SESR-Eval replication package [2].

| LLM (selection criterion) | Confusion matrix | | | | Acc | Lost Ev. | MCC | WMCC |
|---|---|---|---|---|---|---|---|---|
| | TP | FP | FN | TN | | | | |
| Claude 3.7 Sonnet (Acc-best) | 189 | 213 | 326 | 8967 | **94.4%** | 63.3% | 0.387 | 0.466 |
| GPT-4.1 mini (MCC-best) | 289 | 331 | 226 | 8849 | 94.3% | 43.9% | **0.481** | 0.604 |
| GPT-4o (WMCC-best) | 485 | 1743 | 30 | 7437 | 81.7% | 5.8% | 0.401 | **0.724** |
| gpt-4.1-nano (Recall-best, excl. include-all) | 505 | 7598 | 10 | 1582 | 21.5% | 1.9% | 0.093 | 0.228 |
| Ministral 8B (include-all) | 515 | 9180 | 0 | 0 | 5.3% | 0.0% | NaN | NaN |

43.9% of the evidence, whereas the WMCC-best model (GPT-4o, WMCC = 0.724, Recall = 0.942) loses only 5.8%. The 10:1 FN weight in WMCC rewards GPT-4o's vastly higher Recall, correctly reflecting that missing evidence is far more costly than extra screening work. A researcher using MCC alone would select the model that loses nearly half of all relevant papers over one that retains almost all of them. A sensitivity analysis across $w \in [1, 100]$ reveals that the crossover (the $w$ at which the WMCC-best model overtakes the MCC-best) occurs at a median of $w \approx 2.7$ across the 12 disagreement studies (range 1.1–6.4); all crossovers fall below $w = 7$. This means even a modest cost asymmetry — valuing missed evidence just three times more than wasted screening effort — is sufficient to change model selection in most studies, and the default $w = 10$ is comfortably conservative.

3. *Triple metric disagreement demonstrates why MCC alone is insufficient.* Table 5 is, to our knowledge, the first empirical SE example in which Accuracy, MCC, and WMCC each point to a different winner. Accuracy is misleading under class imbalance (as argued throughout this paper); MCC corrects for chance but ignores asymmetric costs and still selects a model that loses 43.9% of evidence; only WMCC integrates chance-correction with cost asymmetry, selecting the model that preserves 94.2% of the evidence.
4. *WMCC is not equivalent to Recall-maximisation.* The same study's `gpt-4.1-nano` has higher Recall (0.981 vs. 0.942) but *lower* WMCC (0.228 vs. 0.724) because it produces 7598 FPs vs. GPT-4o's 1743 (Accuracy 21.5% vs. 81.7%). A pure Recall-maximiser would burden reviewers with ~5,900 additional FPs to recover 0.04 of Recall; WMCC's chance-anchored denominator correctly downgrades classifiers that buy Recall with excessive FPs.

**Why MCC alone is insufficient and how WMCC resolves it: SE evidence from three reanalyses**

*The problem.* MCC is chance-anchored and imbalance-robust, but because it treats false negatives and false positives symmetrically, it can select the wrong classifier or operating point in SR screening. This is not a theoretical concern: in the Felizardo reanalysis, MCC agrees with Accuracy in favouring the threshold that loses more evidence; in the Syriani RL4SE data, MCC selects the classifier that loses 40% of relevant studies over one that loses 18%; and across the SESR-Eval dataset (9 LLMs × 24 SE secondary studies, 34,528 primary studies), MCC and WMCC disagree on the best LLM in 12 of 22 evaluable studies (55%).
*The solution.* WMCC resolves this by encoding the FN:FP cost asymmetry directly into the chance-anchored correlation framework. In every disagreement, the WMCC-best model retains substantially more evidence than the MCC-best model. Sensitivity analysis shows that all ranking flips occur at modest cost ratios (median crossover $w \approx 2.7$, all below $w = 7$), meaning even a moderate preference for retaining evidence over saving screening effort is sufficient to change model selection — and empirically supporting $w = 10$ as a conservative default.

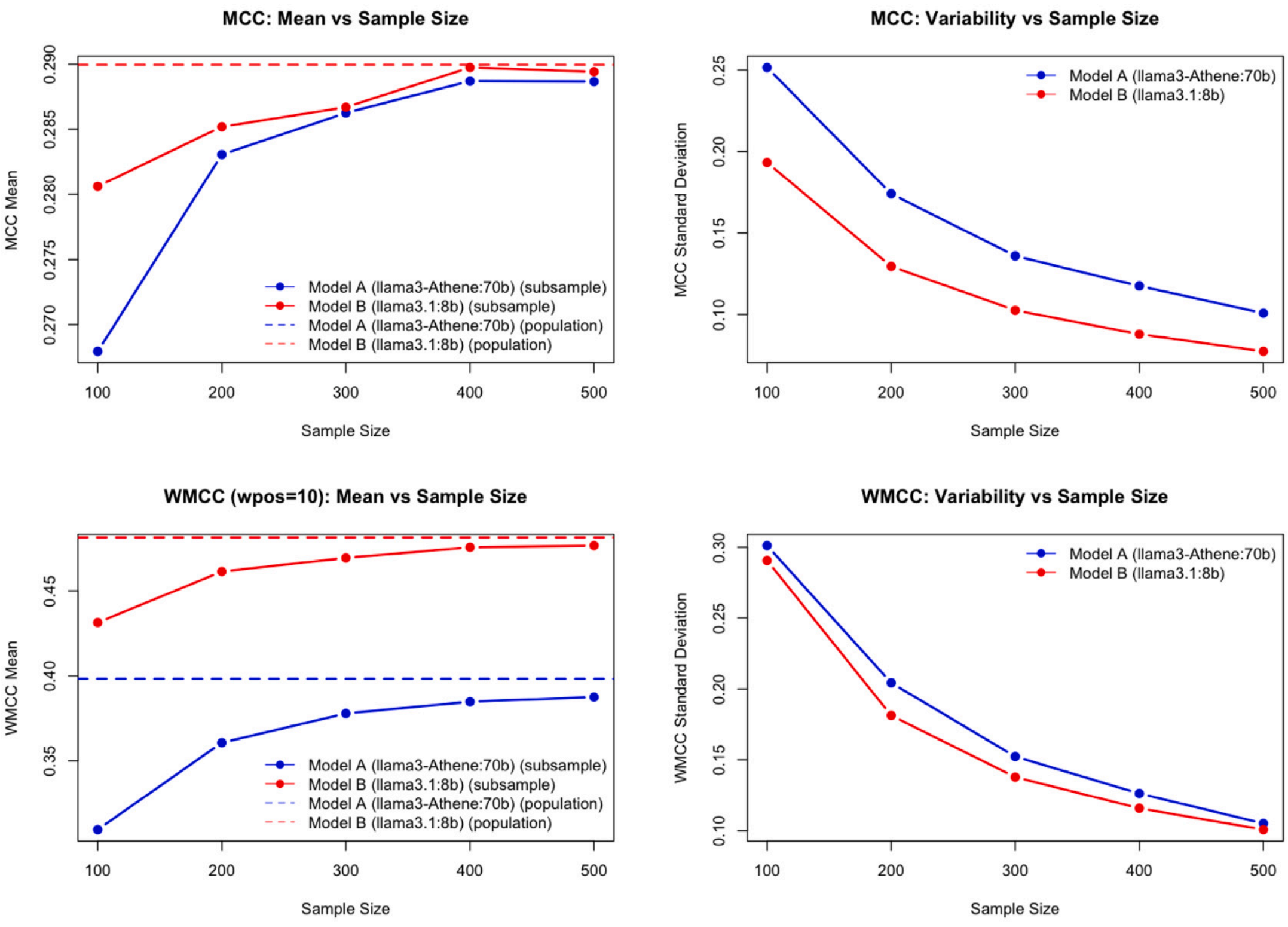

**Fig. 4.** Subsampling stability with 100 to 500 observations.

*Contamination-risk sensitivity analysis.* A concern is that the SESR-Eval patterns could be contamination-driven: the 24 SE secondary studies are publicly indexed and may be in the cloud LLMs' training corpora. As an orthogonal check, we reran the analyses on the complete confusion matrices from DC+ [7] (18 LLMs × 3 biomedical SRs; SI Figs. S1–S3), which uses different domains, different model families (locally served open-weight Llama/Gemma/Mistral/Mixtral/Qwen/Reflection, plus GPT-3.5/GPT-4o), and includes Review II (Neuro) whose underlying SR [46] was published in 2021 — before nearly every tested LLM's training cutoff, making it the *highest*-contamination-risk SR in DC+.

The Accuracy-best LLM differs from the MCC-best LLM in all 3 DC+ SRs and in 26.3% of within-SR LLM pairs (101/384). Pairwise MCC/WMCC ranking flips occur in 31/384 pairs (8.1%), *concentrated in Review II* (28/153, 18.3%). If leakage caused the flip, we would expect the opposite pattern — fewer flips where models had "seen the answer". The flip requires asymmetric error profiles (low FN at the cost of high FP), which leakage tends to erase rather than produce. This is strong evidence the flip is a property of the metrics, not an artefact of contamination. The `llama3.1:8b` vs. `llama3-Athene:70b` comparison in Table 1 (DC+ Review I) is itself one such flip. Full results: `dc_plus_reanalysis.R` in the replication package.

### 3.5. Reporting variation and confidence intervals

The set of papers we reviewed included seven papers that reported measures of variance or confidence intervals, see Fig. 3. However, when we have true and predicted classification for all abstracts related to a specific SR, we have the complete *population* not a sample, so the concept of sampling error is meaningless. Thus, if we report confidence intervals, we need to be very clear to which population and experimental hypotheses they apply.

One situation where CIs are important is when researchers intend to use results of a validation exercise based on a random subset of the available abstracts to decide whether an LLM can be used to analyse the remaining abstracts. In this case, we need to calculate the mean and variance of appropriate performance metrics from the validation sample results. For this purpose, we would suggest using multiple resampling of the validation sample *without replacement*, see the example visualisation in Fig. 4 produced by our R simulation script, in which we reused the known performance metrics of the two models (Model A: llama3-Athene:70b, and Model B: llama3.1.:8b) reported in Table 1.[6] Such visualisations may help to decide whether LLMs can be used to analyse the remaining abstracts, and which models to choose.

It is also worth mentioning that we do not recommend CIs based on the binomial distribution because this assumes that all the data items are independently and identically distributed (iid), which is unlikely for a set of abstracts, and, for the same reason, we do not recommend bootstrapping (i.e., sampling with replacement).

Other situations where CIs are useful include:

1. Assessing whether one LLM generally performs better than another. Here we need to assess the variation between appropriate performance metrics for the LLMs across multiple SRs. This is usually the goal of meta-analysis. It requires analysts to ensure that all metrics are derived from different SRs. A similar approach can be used to assess whether different prompt strategies generally perform better than others.

[6] We developed an R script that generated 10,000 subsample distributions of different sizes (100, 200, 300, 400, 500) for MCC and WMCC that are in line with the performance metrics reported in Table 1 and visualised the results.

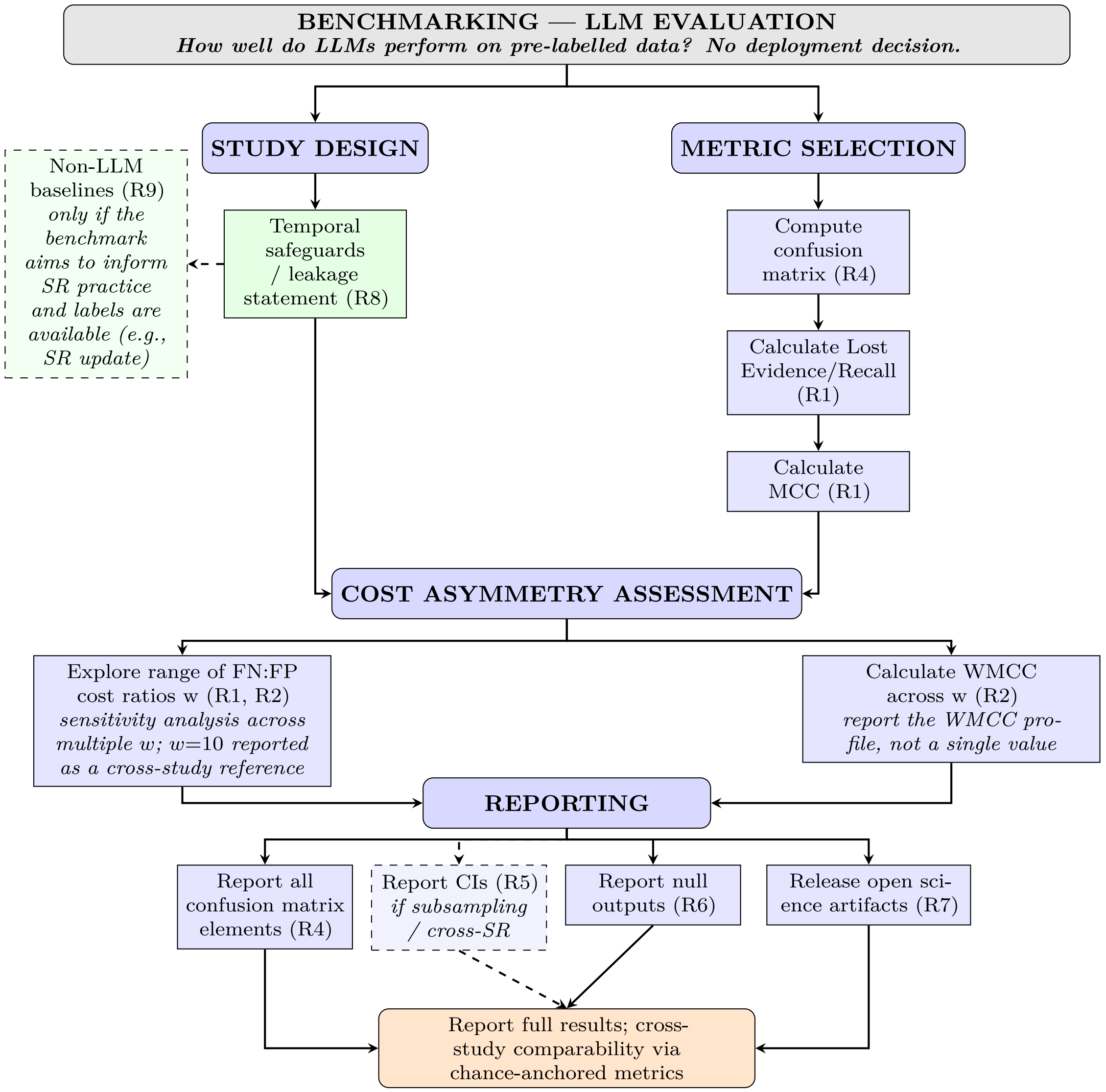

**Fig. 5.** Decision tree for LLM-based SR screening evaluation: **benchmarking studies** (usually retrospective). Dashed elements indicate conditional steps (R5 CIs apply when the benchmark uses subsampling or assesses cross-SR variation, and may be omitted for a single-dataset benchmark that reports the full confusion matrix; R9 applies only when the study design provides training data). Numbers in parentheses refer to the recommendations in the text; R4 appears in both the Compute and Report steps because the same confusion matrix is computed once and then reported in full. Companion: Fig. 6 for deployment.

2. Assessing the extent of LLM inconsistency. For such an analysis, we need individual prompts for the same abstract to be repeated.

However, in either case, it is important to design the evaluation such that repeated analyses of the same prompts or the same abstracts in order to assess performance metric variability do not lead to any data leakage.

### 3.6. Review limitations

A limitation of this investigation is that our review of existing studies is based on a convenience sample of primary studies. We sourced papers from our own knowledge, but in order to address the criticism that we might have explicitly selected papers that support our own opinions, we also included all the primary studies included in two systematic reviews [15,16]. These SRs were the only ones we were able to find, and they were not published by SE-related journal/conferences. However, Kim et al. [15] did find one of the five SE related papers we found, so clearly SE-related papers had not been excluded. Another limitation that applies to any SR is that it includes only previously published papers and searches would have been conducted well before the SR's publication date. However, both SRs provided selections of screening papers obtained independently of the papers we found by our informal searches. Furthermore, good and bad practices were found in papers from all three sources again suggesting that our set of papers were not biased. Although not biased, the simple fact that GenAI is a hot topic means that as soon as a set of published papers is collated and analysed, it will be out of date. This means our list of good and bad practices may be incomplete. In addition, we cannot predict the rate of adoption of various good and bad practices from the analyses presented in this paper.

Another limitation is that the data extraction was performed by one researcher (Kitchenham). However, most of the extracted data were objective and easy to find (or confirm were missing). The most subjective element was deciding whether or not the paper recognised the difference between FPs and FNs in the context of literature screening. Some papers explicitly reported that FNs were more important than FPs, while others reported that Recall was the most important performance metric without explaining why.

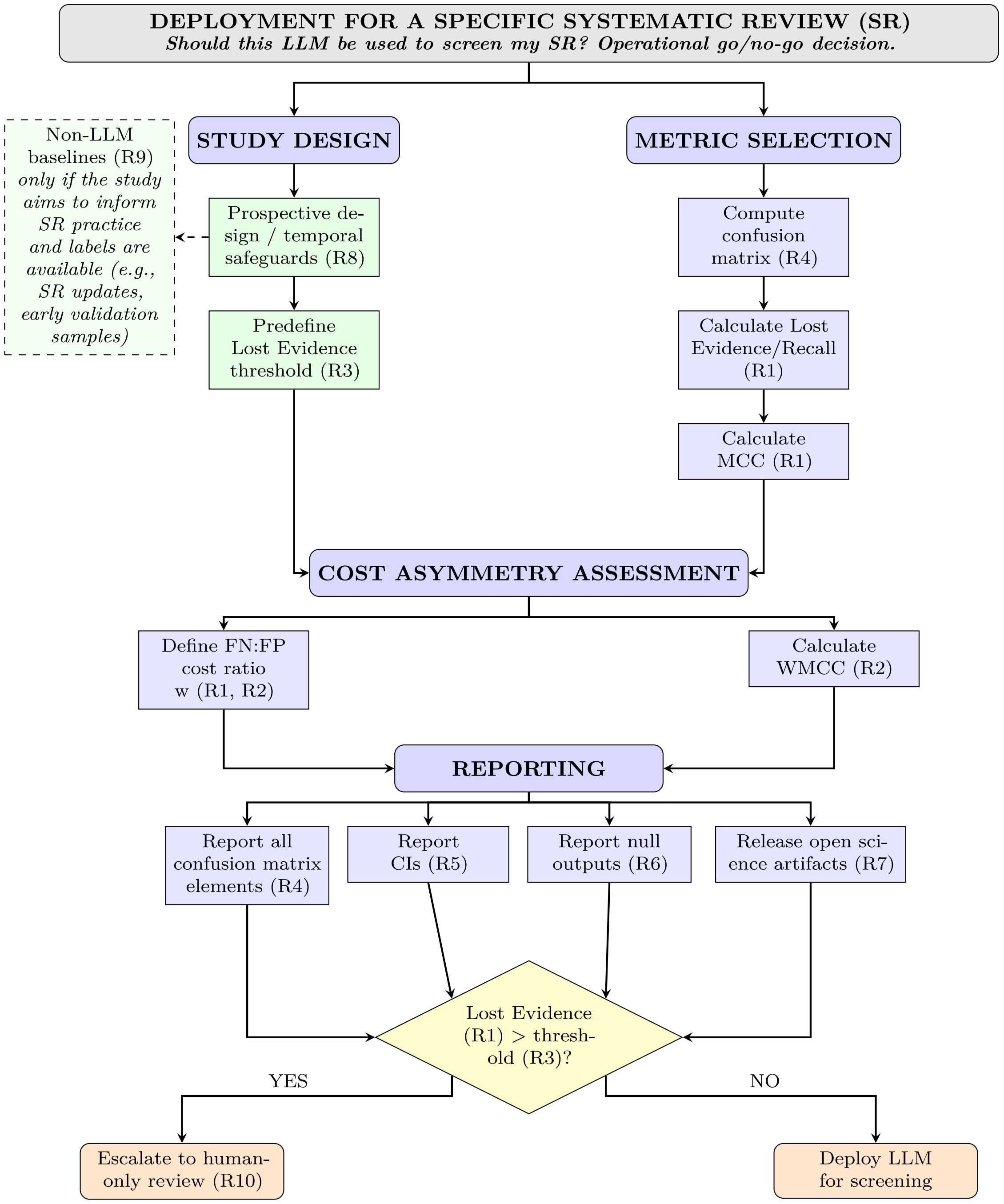

**Fig. 6.** Decision tree for LLM-based SR screening evaluation: **deployment** for a specific SR. Dashed elements indicate conditional steps (R9 applies primarily in SR updates where prior screening provides training data). The decision point checks whether Lost Evidence exceeds the pre-specified threshold (R3), leading to escalation (R10) or deployment. R4 appears in both the Compute and Report steps because the same confusion matrix is computed once and then reported in full. Companion: Fig. 5 for benchmarking.

Finally, our recommendations were developed for title-and-abstract screening. Section 4.3 discusses their principled extension to full-text screening and data extraction based on the non-SE evidence in Cao et al. [28] and Khraisha et al. [9], but this extension remains largely theoretical: no SE-specific empirical worked example yet exists for full-text screening or data extraction, so the transferability claims in Table 6 should be read as hypotheses to be tested rather than validated results. Empirical validation of WMCC on SE full-text screening and on the multi-class or ordinal extraction tasks typical of SE (mapping-study classification, quantitative extraction, quality assessment) is important future work.

## 4. Discussion and conclusions

In this paper, we have presented LLM4SCREENLIT — a set of recommendations for evaluating LLMs for literature screening in SRs — motivated by some critical challenges we have demonstrated. We believe performance metrics problems arise because researchers do not always select metrics that are mapped to the actual needs of the problem domain. A better approach is to report the individual confusion matrix counts together with a small number of performance metrics focussed on the specific research questions being investigated.

In particular, we propose that evaluations of literature screening must prioritise chance-anchored metrics such as MCC and, additionally, explicitly reflect FN vs. FP asymmetry via cost-sensitive analysis. For this purpose, we propose Weighted MCC (WMCC) as a principled extension that retains MCC's correlation meaning and robustness to class imbalance, while addressing the challenge of asymmetric misclassification costs by encoding domain-specific cost ratios. We note that WMCC is not a replacement for MCC. For meta-analysis purposes, MCC allows researchers to assess the predictive capability of different LLMs across different SRs, irrespective of their choice of WMCC weight.

While MCC and WMCC summarise performance at a single operating point, we acknowledge that reporting performance across multiple operating points would strengthen evaluation studies when continuous LLM outputs (e.g., confidence scores) are available. In such cases, researchers could report Lost Evidence–workload curves, decision curve analysis, or performance at multiple Lost Evidence levels to help practitioners understand the Lost Evidence–workload trade-off. The specific levels should be determined by the domain and SR type. However, most current LLM screening studies use binary prompts (include/exclude) that naturally produce single operating points rather than tunable thresholds.

### 4.1. Implications

There are many implications stemming from this paper for researchers (i.e., individuals studying the capabilities of LLMs) and practitioners (i.e., individuals validating GenAI tools as part of the conduct of a specific SR), as well as for those responsible for setting journal and conference policies. Researchers should prioritise prospective, leakage-aware benchmarks and standardise WMCC reporting to enable credible conclusions and MCC for robust meta-analytic synthesis across SRs. They also need to be aware of the difference between the studies based on previously published SRs and studies undertaken to validate LLMs in the context of conducting a new SR. It is important that they address the implications of their results for practitioners in terms of how their results can support SR-based validation exercises. The implications differ by study type (see the benchmarking vs. deployment distinction in Section 4.2). For researchers conducting retrospective benchmarks, the core requirements are complete confusion matrices, chance-anchored metrics (MCC, WMCC), and leakage safeguards. Practitioners deploying LLMs on a specific SR should adopt a reporting kit of their LLM validation exercises that includes complete confusion matrices based on a validation random sample, and performance metrics including Lost Evidence (or Recall), and MCC with confidence intervals. Thresholds for maximum Lost Evidence should be set in advance and are context, domain, and SR type dependent, reflecting the risk tolerance and objectives of the specific review, and should be determined by stakeholders based on the specific consequences of missed evidence. This means that Lost Evidence from FNs is bounded by design. Decisions can then be made based on which (if any) classification processes (i.e., specific combination of one or more LLM(s) with either one or zero human researchers) achieved acceptable values of Lost Evidence (e.g. Lower Confidence interval >0.8 and produced a genuine prediction (i.e., delivered a 95% Lower confidence value of MCC > 0). Any classification processes that achieved these criteria would then be ranked based on WMCC values, with the process with the best WMCC being selected (assuming the WMCC values were greater than zero). If preliminary validation of LLMs supported classification process suggests that performance cannot meet or exceed acceptable levels for a specific SR task, and there are no clear indications of how prompts could be usefully refined, the task must be performed by human researchers.

A major outstanding issue is how the relative costs of FPs and FNs can be determined. Whether we use WMCC (or any other form cost/benefit assessment), we need a weighting factor based on the relative costs of FNs and FPs. In this study, we used 10:1 which might be acceptable for the SE domain, but we have no independent rationale for this value. For commissioned SRs, it may be possible to ask the commissioning group or other stakeholder groups for input about the issue. Assessing relative costs is also made more complex by the fact that not all papers are of equal importance, missing a study with serious methodological flaws is not as important as missing a rigorous study. This discussion also suggests that we need information about the rate of studies that are currently missed by the current gold standard process of screening by two (competent) researchers with disagreements being resolved. Such information would at least provide some information about current levels of FNs.

Journals, conferences, and SR guidelines should require confusion matrices, uncertainty estimates (when appropriate), baseline comparators, explicit leakage/contamination statements, and open artefacts (prompts, seeds, and any materials/artefacts), while discouraging accuracy-focused reporting.

### 4.2. Recommendations

Although, in Sections 2.5 and 3.2, we have discussed a variety of good practices for evaluating LLM screening performance, our main goal was to provide the LLM4SCREENLIT recommendations — actionable guidance for (i) Researchers and practitioners, as well as (ii) Policymakers (e.g., journals, conferences, guideline authors) on how to deal with the observed challenges associated with evaluating the performance of LLMs for screening literature for SRs. In addition to focusing our recommendations on specific target audiences, we decided to organise recommendations into decision-centric themes to improve comprehension. To provide practical guidance, Figs. 5 and 6 present two decision trees — one for benchmarking studies and one for deployment studies — that illustrate the workflows for evaluating LLMs for SR screening. Each tree is organised into two parallel tracks: (1) Study Design steps to complete before running the evaluation, and (2) Metric Selection steps to apply during and after evaluation. These tracks converge at a cost assessment phase and a reporting phase (which explicitly includes reporting all confusion matrix elements). The benchmarking tree (Fig. 5) terminates with a reporting endpoint, while the deployment tree (Fig. 6) adds a decision point based on the predefined Lost Evidence threshold, leading to either deployment or escalation to human review.

Before presenting the recommendations, we distinguish between two study types that share most of the evaluation mechanisms but differ in their purposes: benchmarking studies and deployment studies. *LLM benchmarking studies* usually evaluate one or more LLMs on previously completed SRs, in order to compare models, prompts, or datasets. Their key question is "*How well do LLMs perform at SR screening?*" *LLM deployment studies* are usually prospective. Typically, they test LLMs on a random sample of candidate primary studies from an SR that is currently being conducted, in order to decide whether to deploy the LLM on the remaining (unscreened) abstracts. Their key question is "*Should this LLM be used to screen my SR?*". Because they lead to an operational decision, they need a pre-specified Lost Evidence threshold, confidence intervals from the sample, and an escalation rule if the threshold is not met. We tag each recommendation R1–R10 below with the study type(s) to which it applies.

As a result, we have the following recommendations organised by target audience and themes:

## Target Audience: Researchers and Practitioners

#### Metrics and cost-sensitive evaluation

**(R1) Standardise reporting on Lost Evidence (Recall), MCC, and Weighted MCC (WMCC) with explicit justification of FN:FP cost ratios, and report only relevant metrics while avoiding Accuracy/PABAK as primary metrics** *[Both study types]* (origin: (P9) in Section 2.5 & Section 2.3).

**(R2) Base comparative conclusions on cost-sensitive analyses that reflect asymmetric misclassification costs, using WMCC to combine chance-correction with cost asymmetry and avoiding over-optimising Recall alone** *[Both study types]* (origin: (P9) in Section 2.5). For benchmarking studies where domain-specific cost ratios are not available, report a sensitivity analysis across multiple $w$ values as the primary result; $w = 10$ can additionally be reported as a cross-study reference point for comparability with deployment studies.

**(R3[*]) When deploying an LLM for a specific SR, predefine acceptable Lost Evidence (minimum Recall) thresholds as guardrails for the review's risk tolerance and objectives, aligned to review type (e.g., SR, Mapping/Scoping Study, Rapid Reviews) and domain (e.g., healthcare, software engineering)** *[Deployment only]* (origin: Section 4.1).

#### Reporting and transparency

**(R4) Publish complete confusion matrices for every model, dataset, and prompt to enable recomputation of necessary metrics like Lost Evidence, MCC, WMCC, and alternative (e.g., cost–benefit) analyses and future meta-analyses**[a] *[Both study types]* (origin: (P1) in Section 2.5).

**(R5) For validation tests based on samples, report uncertainty for each performance metric via confidence intervals and document the estimation method used. When testing LLMs on a random sample of abstracts to decide whether to deploy them on the remaining abstracts, use resampling without replacement to estimate confidence intervals for likely performance on the full dataset** *[Conditional — see text]* (origin: (P3) in Section 3.2; resampling method proposed by the authors in Section 3.5 and illustrated in Fig. 4). For benchmarking studies, the deployment-specific resampling guidance (second clause) does not apply, but CIs remain relevant when the benchmark uses subsampling or assesses cross-SR variation (see Section 3.5).

**(R6) Quantify LLM output consistency and null or invalid outputs, specify the evaluations rule that treats unclassifiable or referred-back items for fair metric computation** *[Both study types]* (origin: (P4) and (P5) in Section 3.2).

* An asterisk '*' means that the recommendation is optional.
[a] DC+ revealed a consistent problem with lost evidence across three reviews and 18 LLMs. This important result is only clear because they reported all their confusion matrices.

**(R7) Release open science artefacts, including prompts, seeds, code, and curated data, to support open science and independent verification** *[Both study types]* (origin: (P2) in Section 2.5).

#### Study design and validity

**(R8) Use designs that prevent training–test overlap: prospective data collection for deployment studies, or documented temporal/contamination safeguards for retrospective benchmarks; explicitly state the leakage risks and the mitigations applied** *[Both study types]* (origin: (P7) & (P8) in Section 3.2).

**(R9) Consider non-LLM baselines when the study aims to inform SR practice (e.g., deployment-situated studies, SR updates) and the design provides labels for training baselines. Pure cross-LLM benchmarks and prompting-strategy evaluations satisfy neither condition and need not include non-LLM baselines** *[Conditional and optional — see text]* (origin: (P2) in Section 2.5).

#### Decision thresholds and operations

**(R10) When observed Lost Evidence exceeds the prespecified threshold, escalate to human review or adjust prompts/models to maintain SR validity** *[Deployment only]* (origin: Section 4.1).

## Summary: Applicability by Study Type

The table below summarises which of R1–R10 apply to each study type. Benchmarking studies do not need R3 and R10 because they do not make deployment decisions; R5's deployment-specific resampling guidance does not apply, but reporting CIs remains relevant when benchmarks use subsampling or assess cross-SR variation. R9 is conditional and optional in both tracks — optional in the strict sense that researchers are advised to *consider* non-LLM baselines rather than being required to include them.

#### Benchmarking studies (usually retrospective, no deployment decision)

**Required:** R1 (metrics: Lost Evidence, MCC, WMCC), R2 (cost-sensitive; sensitivity analysis across multiple $w$ values is the primary result, with $w = 10$ reported additionally as a cross-study reference point), R4 (full confusion matrices), R6 (consistency & null outputs), R7 (open artefacts), R8 (leakage safeguards).

**Conditional:** R5 (CIs) — the deployment-specific resampling guidance does not apply, but CIs are relevant when benchmarks assess cross-SR variation or use subsampling (see Section 3.5); R9 (non-LLM baselines) — not required when the benchmark's sole aim is cross-LLM comparison or prompting-strategy evaluation; consider including baselines only if the benchmark additionally aims to inform operational SR use (e.g., SR updates, reused labelled datasets from which baselines can be trained).

**Not applicable:** R3 (predefined Lost Evidence threshold), R10 (escalation to human review).

**Deployment for a specific SR (prospective, validation)**

**Required:** R1, R2, R3, R4, R5, R6, R7, R8, R10.

**Conditional:** R9 — consider non-LLM baselines when the study aims to inform SR practice and a labelled sample is available (from prior screening in an SR update, or from early validation in a new SR, cf. [41]).

**Target Audience: Policymakers (Journals, Conferences, Guideline authors)**

**Metrics and cost-sensitive evaluation**

**($R1_{PM}$) Require reporting of Lost Evidence (Recall), MCC, and WMCC with declared FN:FP cost ratios, and discourage accuracy-centric or PABAK-focused reporting as primary evidence** (origin: Section 4.1).

**($R2_{PM}$) Mandate cost-sensitive evaluation narratives that explain trade-offs between efficiency and Lost Evidence, referencing WMCC or equivalent methods** (origin: Section 4.1).

**Reporting and transparency**

**($R3_{PM}$) Require complete confusion matrices for all reported metrics to enable recomputation and meta-analytic synthesis** (origin: Section 4.1).

**($R4_{PM}$) Require disclosure of LLM output consistency and null-output rates, with explicit rules for handling unclassifiable or referred-back items in evaluation** (origin: Sections 2.4 and 4.1).

**($R5_{PM}$) Require open artefacts (prompts, seeds, code, data, and materials) to support independent verification and reproducibility** (origin: Section 4.1 and (P2) in Section 2.5).

**Study design and validity**

**($R6_{PM}$) Require explicit leakage/contamination statements and temporal or provenance safeguards in retrospective or benchmark-based studies** (origin: Section 4.1, (P7) and (P8) in Section 3.2).

**($R7_{PM}$) When the study design permits (e.g., SR updates, retrospective benchmarks with labelled datasets), encourage inclusion of non-LLM baselines for claims about efficiency or effectiveness; disallow claims based on accuracy-only evidence** (origin: Section 4.1).

**Decision thresholds and governance**

**($R8_{PM}$) Encourage pre-registration of acceptable Lost Evidence (minimum Recall) thresholds and escalation rules as part of protocol submissions, aligned to domain risk** (origin: Section 4.1).

### *4.3. Extending the recommendations to later SR stages*

Our recommendations were derived from evidence on title/abstract screening, but the underlying principles — class imbalance, asymmetric misclassification costs, and the need for chance-anchored metrics — are mathematical rather than stage-specific. Two of the papers in our review explicitly address later stages: Cao et al. [28] evaluate both abstract and full-text screening using prompt engineering strategies across multiple LLMs, and Khraisha et al. [9] report performance at three stages (title/abstract screening, full-text screening, and data extraction) using the same evaluation metrics. Drawing on these two papers and on the nature of each recommendation, Table 6 summarises the applicability of R1–R10 across SR stages.

At the full-text screening stage, the task remains binary (include/exclude), so all recommendations transfer directly. Cao et al. [28] show that prompt engineering strategies (Framework CoT, ISO-ScreenPrompt) that differ from abstract-screening strategies can substantially improve full-text performance, but the evaluation metrics remain the same. Class imbalance is typically less extreme at full-text (both Cao et al. and Khraisha et al. [9] report higher inclusion ratios than for title/abstracts), however the asymmetric FN/FP costs remain.

At the data extraction stage, the task shifts from binary classification to structured information retrieval, so recommendations R1, R2, R3, R4, and R10 require not merely reinterpretation but a different evaluation design. Khraisha et al. [9] use the same metrics (Sensitivity, Specificity, Accuracy, Cohen's kappa) for data extraction as for screening, treating each extraction field as a binary detection problem. While this is a reasonable first approximation, a richer evaluation would report per-field correctness, completeness, and spurious-extraction rates against a human-reference gold standard — an adaptation of R4 (confusion matrix) to the extraction context. Importantly, MCC generalises to multi-class problems via the $K{\times}K$ confusion matrix (see Gorodkin [47]), and WMCC can be naturally extended by the same logic: weighting each cell of the $K{\times}K$ confusion matrix by a corresponding entry in a $K{\times}K$ cost matrix (rather than a single scalar $w$) before applying Gorodkin's formula. The *principles* that motivate these metrics — chance-anchored evaluation and asymmetric misclassification costs — thus carry over to multi-class extraction tasks, although the cost matrix must be elicited from domain stakeholders for each specific extraction context.

In the SE domain, data extraction tasks are frequently heterogeneous: mapping studies often involve multi-class categorisation of primary studies along facets such as research method, SE domain, contribution type, and venue type [22], SRs may require extracting quantitative data (effect sizes, sample sizes, and contextual variables), and quality assessment involves applying ordinal checklists rather than binary decisions. These tasks differ substantially from the binary per-field approach used by Khraisha et al. [9] and from biomedical extraction, where fields tend to be more structured (e.g., patient counts, drug doses, specific outcomes).

We emphasise that this applicability analysis is largely theoretical for later stages: the full-text screening row rests on two empirical sources (involving the same binary task), while the data extraction row rests on a single study plus the authors' principled argument. Adapting the evaluation principles from this paper to these diverse SE extraction tasks — particularly the heterogeneous mapping, quantitative, and quality-assessment tasks described above — requires empirical validation that we flag as an open question for future work.

### *4.4. Guidance for editors and reviewers*

The policymaker recommendations ($R1_{PM}$–$R8_{PM}$) specify *what* venues should require; this subsection offers practical guidance on *how* to enforce them.

*Reviewer/editor checklist*

The following checklist can be adopted by SE journals (such as IST) and conferences as part of their author guidelines or submission checklists for LLM-based SR screening studies. Authors should submit a completed copy of this checklist as part of the submission package, confirming which items are fulfilled and declaring any items that are not applicable with a brief justification. Fillable versions of this checklist in four formats (LaTeX, Word, Markdown, plain text), plus a PDF preview, are available in the replication package at https://doi.org/10.6084/m9.figshare.31356613.

**Table 6**
Applicability of R1–R10 across SR stages. T = applies as stated; P = principle applies (the underlying principle carries over but the binary operationalisation requires adaptation — see footnotes); C = conditional (as for title/abstract screening).

| | R1 | R2 | R3 | R4 | R5 | R6 | R7 | R8 | R9 | R10 |
|---|---|---|---|---|---|---|---|---|---|---|
| Title/Abstract (origin)[d] | T | T | T | T | T | T | T | T | C | T |
| Full-text screening[e] | T | T | T | T | T | T | T | T | C | T |
| Data extraction[f] | P[a] | P[a] | P[b] | P[c] | T | T | T | T | C | P[b] |

[a] MCC and WMCC do not directly apply to structured information retrieval; the *principles* (chance-anchored, cost-asymmetric evaluation) transfer but require per-field or per-record metrics.
[b] Risk tolerance and escalation principles carry over, but thresholds must be defined per-field rather than per-study.
[c] The analogue of a confusion matrix is a per-field accuracy table (correct, missing, and spurious extractions vs. a human-reference gold standard).
*Evidence basis:*
[d] 29 reviewed papers.
[e] Cao et al. [28] and Khraisha et al. [9] (2 non-SE studies); task remains binary.
[f] Khraisha et al. [9] (single non-SE study) + authors' principled argument; SE-specific validation needed.

**Table 7**
Empirical performance categories for MCC and WMCC in SE LLM screening, derived from terciles of 183 evaluable configurations in the SESR-Eval reanalysis (9 LLMs × 24 SE secondary studies), following the approach of Kampenes et al. [48]. Medians are the median values within each tercile. These categories reflect the current state of LLM screening technology (2025 snapshot) and will shift as LLMs improve.

| | Small (lower tercile) | | Medium (middle tercile) | | Large (upper tercile) | |
|---|---|---|---|---|---|---|
| | Range | Median | Range | Median | Range | Median |
| MCC | <0.21 | 0.09 | 0.21–0.38 | 0.31 | >0.38 | 0.46 |
| WMCC ($w = 10$) | <0.19 | 0.07 | 0.19–0.42 | 0.30 | >0.42 | 0.56 |

**LLM4SCREENLIT Reviewer/Editor Checklist**

- ☐ Complete confusion matrix (TP/FP/TN/FN) for every model × SR × prompt (R4, $R3_{PM}$).
- ☐ Lost Evidence (1 − Recall) reported and discussed (R1, $R1_{PM}$).
- ☐ MCC reported unrescaled in $[-1, 1]$ (R1, $R1_{PM}$).
- ☐ WMCC with explicit weight $w$ and justification, even if $w = 1$ (R1, R2, $R2_{PM}$).
- ☐ Null/invalid output rate, suspected cause, and handling rule specified (R6, $R4_{PM}$).
- ☐ Leakage/data-contamination statement with mitigation (R8, $R6_{PM}$).
- ☐ Replication package with prompts, seeds, code, and labelled data (R7, $R5_{PM}$).
- ☐ *Deployment studies only:* pre-specified Lost Evidence threshold with domain-stakeholder justification (R3) and escalation rule (R10, $R8_{PM}$).
- ☐ *Benchmarking studies only:* statement that deployment decisions are out of scope; R3/R10 are not required.
- ☐ *When the study aims to inform SR practice and training labels are available:* non-LLM baseline comparison included (R9, $R7_{PM}$).

*Interpreting the metrics*

To help reviewers calibrate expectations, we derive empirical performance categories from the 183 evaluable LLM×study configurations in the SESR-Eval reanalysis (9 LLMs, 24 SE secondary studies; Table 5). Following Kampenes et al. [48], who defined empirical small/medium/large effect-size categories for SE experiments by splitting observed values into terciles (lower 33%, middle 34%, upper 33%) and reporting the median within each category, we apply the same approach to MCC and WMCC. Table 7 presents the resulting performance categories for current LLM screening performance in SE.

Approximately 8% of configurations had negative MCC values (worse than chance-level performance). Reviewers should note that WMCC < MCC is expected when $w > 1$ (the FN penalty pulls the score down), and that WMCC > MCC signals a classifier with unusually high Recall relative to its overall correlation. These categories should be treated as empirical reference points, not permanent thresholds (see Table 7 caption).

*Minimum reporting template*

To illustrate what a minimally compliant results table should contain, Table 8 shows a template that can be adapted by authors. Each row represents one model–prompt–SR combination; the columns cover the confusion-matrix counts and the metrics recommended in R1 and R2. Authors should provide one such table (or equivalent) for every evaluated configuration, together with the FN:FP cost ratio $w$ and any confidence intervals (R5) when sample-based validation was used.

For worked examples of compliant reporting, see Table 3 (Felizardo et al. reanalysis: four model×threshold configurations with all recommended columns), Table 4 (Syriani et al. reanalysis: LLM vs. non-LLM baselines across five SE datasets), and Table 5 (SESR-Eval reanalysis: metric disagreement across nine LLMs). Ready-to-use R and Python functions for computing all recommended metrics from confusion-matrix counts (`llm4screenlit_metrics.R` and `llm4screenlit_metrics.py`) are included in the replication package at https://doi.org/10.6084/m9.figshare.31356613.

*Enforcement actions*

Concrete steps venues can take to operationalise the policymaker recommendations:

1. Strongly recommend (or even require) authors to submit a completed compliance declaration based on the checklist above, confirming that each applicable item is fulfilled and justifying any items marked as not applicable. Highlight this on the journal's submission guidelines web page. This follows the practice widely adopted by medical journals, which require authors to submit completed PRISMA [45] checklists alongside their manuscripts. In the SE domain, SEGRESS [22] provides an analogous reporting checklist for secondary studies; our checklist extends this approach to LLM-based screening evaluations specifically.

**Table 8**
Minimum reporting template for an LLM-SR screening evaluation. Authors fill one row per model × prompt × SR combination. Italicised columns are deployment-only (R3, R5). The grey row shows a filled example from Table 4.

| Model | Prompt | SR | TP | FP | FN | TN | Recall | Lost Ev. | MCC | WMCC | Null% | *CI* |
|---|---|---|---|---|---|---|---|---|---|---|---|---|
| GPT-3.5-turbo-0613 | PositiveX | MobileMDE | 47 | 50 | 8 | 187 | 0.855 | 14.5% | 0.534 | 0.624 | 1.7% | – |
| … | … | … | … | … | … | … | … | … | … | … | … | … |

Study type: Benchmarking/Deployment (delete improper). FN:FP cost ratio $w$ = ____ (justification: ____________).

2. Require complete confusion matrices as supplementary material at submission time ($R3_{PM}$). Recommend that results tables follow the minimum reporting template (Table 8) or an equivalent format covering all recommended columns.
3. Reject or require revision for submissions that lack any chance-anchored metric and report only Accuracy or PABAK as their primary classification performance measure ($R1_{PM}$).
4. Require an explicit FN:FP cost-ratio declaration ($w$) in the method section; $w = 1$ is acceptable only with justification ($R2_{PM}$).
5. Require a leakage/contamination statement in retrospective or benchmark studies; flag papers using public benchmarks without one ($R6_{PM}$). An example statement: *"The SRs used for evaluation were published in [year range], [before/after] the training data cutoff of [LLM version]. We [did/did not] verify that the screened abstracts were not in the LLM's training corpus. The risk of data contamination is [low/moderate/high] because [justification]"*.
6. Require a replication-package URL at submission ($R5_{PM}$).
7. Adopt the checklist above as part of the venue's author guidelines. Ready-to-use policy text (short and extended versions) for inclusion in author guidelines or calls for papers is available in the replication package at https://doi.org/10.6084/m9.figshare.31356613.

### *4.5. Future research*

The LLM4SCREENLIT recommendations consolidate nine good practices (see Sections 2.5 and 3.2) observed across the reviewed literature, demonstrate the pitfalls of accuracy-centric reporting under class imbalance, and propose WMCC to integrate chance-correction with cost asymmetry, thereby turning fragmented results into operational guidance for SR screening, offering a coherent evaluation framework that supports more credible decisions. In spite of this, further research should investigate the effectiveness of combining multiple good practices identified by us (P1–P9) with ones identified by other researchers to keep the evaluation framework as robust as possible. Also, retrospective studies of past SRs could include the use of additional human researchers performing the classification process to identify the actual performance of the current screening practice (i.e, two humans with disagreements resolved), as well as assessing the performance of using human–AI teams to generate classifications. See, for example, Cao et al. [28] who, although having data from several different completed SRs, based their study of prompt engineering method around samples from the different SR datasets and employed four human researchers to perform classifications on the samples to provide a rigorous assessment of both human and AI tool performance.

## CRediT authorship contribution statement

**Lech Madeyski:** Writing – review & editing, Writing – original draft, Visualization, Software, Methodology, Investigation, Formal analysis, Data curation, Conceptualization. **Barbara Kitchenham:** Writing – review & editing, Validation, Methodology, Investigation, Conceptualization. **Martin Shepperd:** Writing – review & editing, Software, Methodology, Formal analysis, Conceptualization.

## Declaration of competing interest

The authors declare that they have no known competing financial interests or personal relationships that could have appeared to influence the work reported in this paper.

## Acknowledgements

We thank the anonymous reviewers for their thoughtful and constructive feedback, which led to significant improvements in the clarity, scope, and presentation of this paper.

## Appendix A

This section reports the formulas used to calculate the metrics discussed in Section 2. The formulas are based on counts obtained from a confusion matrix as shown in Table 9.

Accuracy measures the proportion of all items that are correctly classified:

$$Accuracy = \frac{TP + TN}{TN + TP + FN + FP} \quad (9)$$

Recall, which is also referred to as Sensitivity, measures the proportion of all positives correctly classified

$$Recall = \frac{TP}{TP + FN} \quad (10)$$

Precision measures proportion of all items that were classified as positive that were correctly classified.

$$Precision = \frac{TP}{TP + FP} \quad (11)$$

Specificity, which is also referred to as the True Negative rate, measures the proportion of all negatives that were correctly classified:

$$Specificity = \frac{TN}{TN + FP} \quad (12)$$

F1 is a confusion matrix metric designed to assess retrieval from search engine queries, where the number of true negatives (TNs) cannot be counted:

$$F1 = \frac{2 \times TP}{2 \times TP + FP + FN} \quad (13)$$

PABAK, Prevalence Adjusted Bias Adjusted Kappa [18,19], is defined as:

$$PABAK = 2 \times p_o - 1 \quad (14)$$

where $p_o$ is the observed agreement i.e., the proportion of identical classifications, also known as Accuracy. So if Accuracy = 1, PABAK = 1, if Accuracy = 0, PABAK = −1 and if Accuracy = 0.5 PABAK = 0. This means that PABAK is simply a centred version of Accuracy, and is just as unreliable as Accuracy for imbalanced datasets.

**Table 9**
A confusion matrix based on the classifications assumed to be true and the classifications produced by the prediction model.

| | Gold standard True | Gold standard False | Total |
|---|---|---|---|
| Predicted True | TP | FP | TP+FP |
| Predicted False | FN | TN | FN+TN |
| Total | TP+FN | FP+TN | N |

The Matthews Correlation Coefficient (MCC) is a form of correlation coefficient calculated as:

$$\mathrm{MCC} = \frac{(TP \times TN) - (FP \times FN)}{\sqrt{(TP+FP)\times(TP+FN)\times(TN+FP)\times(TN+FN)}} \quad (15)$$

## Appendix B. Supplementary data

Supplementary material related to this article can be found online at https://doi.org/10.1016/j.infsof.2026.108204.

## Data availability

The replication package (extracted data, analysis scripts, and documentation) is available at https://doi.org/10.6084/m9.figshare.31356613.